\documentclass[aps,prl,10pt,twocolumn,superscriptaddress]{revtex4-2}
\usepackage{amsmath,amssymb,amsfonts}
\usepackage{array}
\usepackage{booktabs}
\usepackage{graphicx}
\usepackage{xcolor}
\usepackage[caption=false]{subfig}
\usepackage[export]{adjustbox}
\usepackage[colorlinks]{hyperref}
\usepackage{bookmark}

\graphicspath{{./images/}}
\newcommand{\ket}[1]{|{#1}\rangle}

\newcommand{\mel}[3]{\langle{#1}|{#2}|{#3}\rangle}
\newcommand{\iu}{\mathrm{i}\mkern1mu}

\renewcommand{\Re}{\operatorname{Re}}
\renewcommand{\Im}{\operatorname{Im}}
\DeclareMathOperator{\Tr}{Tr}
\let\Delta\varDelta

\begin{document}

\title{Virasoro and Kac--Moody algebra in generic tensor network representations of 2d critical lattice partition functions}
\author{Ruoshui Wang}
\affiliation{Cornell University, Ithaca, New York, 14853 USA}
\author{Xiangdong Zeng, Ce Shen}
\affiliation{State Key Laboratory of Surface Physics, Fudan University, 200433 Shanghai, China}
\affiliation{Shanghai Qi Zhi Institute, 41st Floor, AI Tower, No. 701 Yunjin Road, Xuhui District, Shanghai, 200232, China}
\affiliation{Department of Physics and Center for Field Theory and Particle Physics, Fudan University, Shanghai 200433, China}
\author{Ling-Yan Hung}
\email{lyhung@fudan.edu.cn}
\affiliation{State Key Laboratory of Surface Physics, Fudan University, 200433 Shanghai, China}
\affiliation{Shanghai Qi Zhi Institute, 41st Floor, AI Tower, No. 701 Yunjin Road, Xuhui District, Shanghai, 200232, China}
\affiliation{Department of Physics and Center for Field Theory and Particle Physics, Fudan University, Shanghai 200433, China}
\affiliation{YMSC, Tsinghua University, Beijing, China}
\date{\today}

\begin{abstract}
In this paper, we propose a general implementation of the Virasoro generators and Kac--Moody currents in generic tensor network
representations of 2-dimensional critical lattice models. Our proposal works even when a quantum Hamiltonian of the lattice model is
not available, which is the case in many numerical computations involving numerical blockings.
We tested our proposal on the 2d Ising model, and also the dimer model, which works to high accuracy even with a fairly small system size.
Our method makes use of eigenstates of a small cylinder
to generate descendant states in a larger cylinder, suggesting some intricate algebraic relations between lattice of different sizes.
\end{abstract}

\maketitle

\bookmarksetup{startatroot}

\section{Introduction}

Critical lattice models provide a profound insight into the emergence of continuous field theory in a discrete world.
There are families of classical lattice partition functions that reproduce CFT data, most notably the series of minimal models.
These lattice models often are integrable, and their topological/categorical symmetries can be built in in many novel constructions developed recently based on fusion categories~\cite{Frank1, Aasen:2020jwb, frank_haagerup, Tachikawa_Haagerup}. These classical lattice partition functions generically
admit tensor network representations. The framework of tensor networks is a powerful tool for recovering the CFT. Particularly, it is found that the process of blocking and RG flow can be done in a controlled and efficient way with arbitrary accuracy depending on the bond dimension of the tensors kept. Tensor Renormalization Group (TRG)~\cite{Levin_2007} or Tensor Network Renormalization (TNR)~\cite{Evenbly_2015} are examples of these blocking algorithms. It is demonstrated that following a recursive blocking algorithm acting on these tensors, one can efficiently read off at least the low lying primaries and descendants from the ``fixed-point'' tensors. The set of tools thus promises an alternative route to obtaining and perhaps potentially classifying CFTs by constructing their lattice versions with various built-in (categorical symmetries) and solving them numerically to very high accuracy.

One very important property characterising 1+1\,d CFT is the infinite dimensional Virasoro symmetries and more generally the Kac--Moody symmetries. There have been discussions of how Virasoro~\cite{Koo:1993wz, Pasquier:1989kd, Milsted:2017csn, Zou:2017zce} and Kac--Moody operators~\cite{Wang} can be implemented in lattice models.
However, these constructions depend on the knowledge of quantum Hamiltonian retrieved from the transfer matrices of the classical lattice models by taking the highly inhomogenous limit of the parameters. In a generic tensor network representation, particularly one in which we obtained by extracting the fixed-point of the blocking procedure, the individual tensor is given by some numerical values --- there is no analytical parameter that one can expand in to obtain the quantum Hamiltonian. In cases as such, it is unclear how these Virasoro/Kac--Moody generators could be implemented.

In this paper, we develop a novel and efficient technique to construct these operators and insert them into the partition function in an arbitrary tensor network representation.
It has been observed that conformal eigenstates, primaries or descendants, are eigenstates of the partition functions made of tensors tiling a cylinder. It is known that to obtain accurate conformal data involving larger conformal dimensions, it is necessary that the total bond dimension of input/output legs of the cylinder is sufficiently large (i.e.\ by packing a large number of tensors into the cylinder, each with small bond dimensions, or a couple tensors with large bond dimensions).
In this paper, we demonstrate that one can obtain the stress tensor $T$ and conserved current $J$ as eigenstates of a very small cylinder, and use them to construct Virasoro/Kac--Moody descendants that are eigenstates of a very large cylinder, to surprising accuracy. The fact that it works suggests that there are interesting algebraic relations between eigenstates of cylinders of different sizes. As a numerical technique, this would open the door to many applications.

The organization of the paper is as follows. We will begin with a brief introduction of tensor network representations of partition functions.
Then we will dive into the precise methodology that has been tested on the Ising and dimer model.
We will end with some discussions. Some further technical details are relegated to the appendices.

\section{Tensor network and Partition functions}

Many 1+1\,d CFTs are continuous/critical limits of lattice models. These lattice partition functions admit tensor network realizations.
It is well known that the transfer matrices of these lattice models can be expressed as a tensor network, the most well-known being the Ising model, which will be the main focus of the current paper.

A generic tensor network representation of a partition function thus takes the following form, as depicted in Fig.~\ref{fig:tm}.

\section{Virasoro and Kac--Moody from Eigenstates of a small cylinder}
\begin{figure}[ht]
\centering
\includegraphics[width=0.8\linewidth]{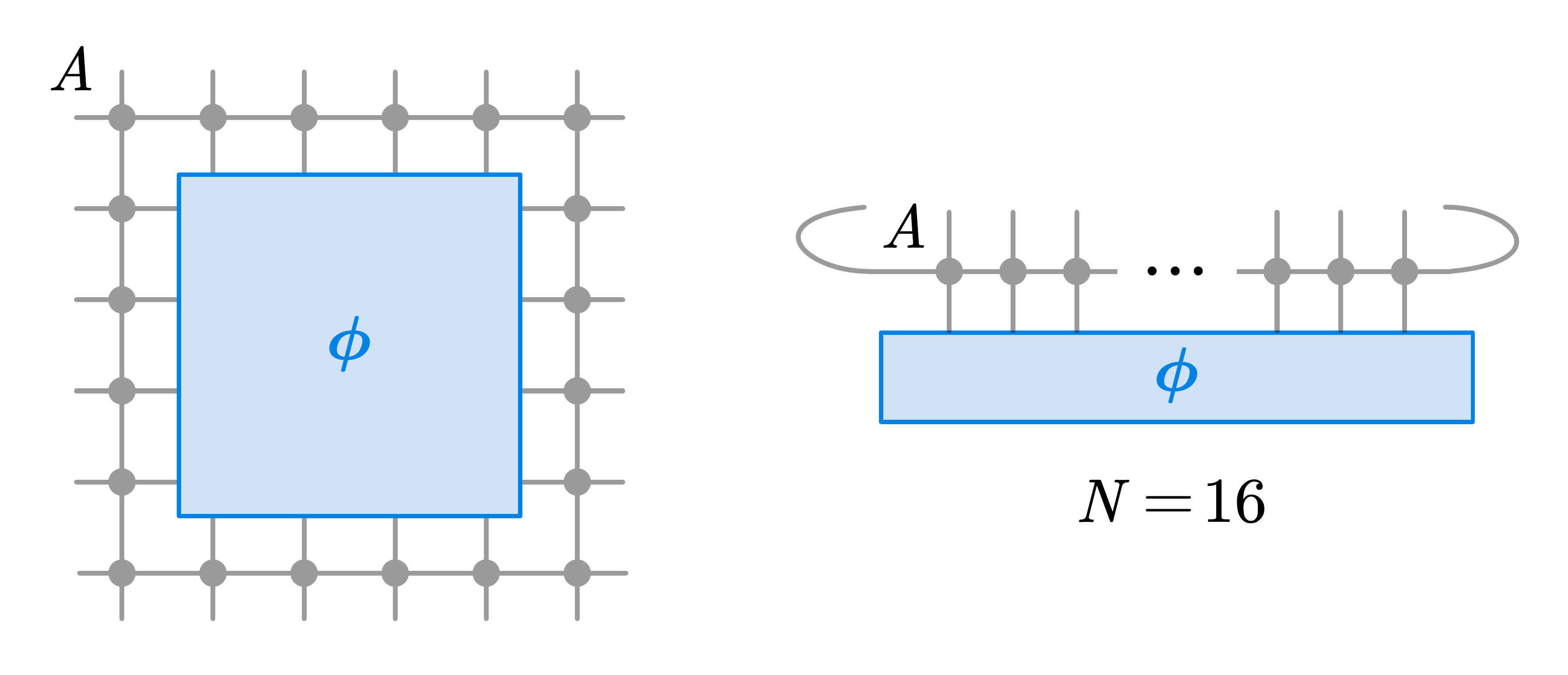}
\caption{(Left) Partition function constructed from four-leg tensors $A$ with an operator $\phi$ inserted. (Right) The tensors $\phi$ corresponding to a conformal operator are obtained by solving for eigenstates of a cylinder built from either $A$ tensors.}
\label{fig:tm}
\end{figure}

In a CFT, it is well known that there is a one-to-one correspondence between states and operators, i.e.\ consider performing a path-integral over a disk with an operator inserted at the origin. This is expected to produce the wavefunction of
a state on the boundary of the disk which corresponds to the inserted operator. If the inserted operator has a well-defined conformal dimension (i.e.\ either a primary or a descendant),
then the state produced would be an eigenstate of the dilation operator. The dilation operator is nothing but the path-integral on an annulus. An annulus with complex coordinates $z$ can be conformally mapped to a path-integral covering a cylinder with coordinates $\omega = \sigma + i \tau$ via the conformal map $z = \exp(\omega)$.
Therefore operator insertions in a planar path-integral can be achieved by cutting out a disk from the plane and replacing the disk by
eigenstates of the cylinder corresponding to the particular conformal operator.
In the case of a lattice partition function, the same local tensors $A_{ijkl}$ defining the lattice partition $Z$ on a plane can be used to cover a cylinder instead.
The cylinder contains one layer of tensors in the vertical direction, and with $N$ sites in the horizontal direction, see also Fig.~\ref{fig:tm}.
It is a matrix with in-coming and out-going indices corresponding to dangling legs of the local tensors at the two boundaries of the cylinder.
Up to normalizations, the eigenstates of this cylinder can be used to approximate operator insertion on a plane. To distinguish the spin of the operators, one can diagonalize
simultaneously with the translation operator $P$ to obtain the finite-size spectrum $\{ \phi_\alpha \}$ along with the conformal data~\cite{Levin_2007, Evenbly_2015, Hauru_2016, Evenbly_2016, Yang_2017, Frank1}, i.e.\ scaling dimensions $\Delta_\alpha$ and the conformal spins $s_\alpha$.

In analogy to the Virasoro generators in the continuum, the lattice version on an $N$-site cylinder can be written as
\begin{equation}
\label{eq:Vir}
L_n \sim \sum_{j=1}^N e^{\iu j n \frac{2\pi}{N}} T(j), \quad \bar{L}_n \sim \sum_{j=1}^N e^{-\iu j n \frac{2\pi}{N}} \bar{T}(j)
\end{equation}
where $T(j)$/$\bar{T}(j)$ are lattice representations of energy-momentum tensor acting on lattice site $j$.

However, for a general tensor network representation of the partition function, the energy-momentum tensor $T$ is often unknown analytically. To overcome this obstacle, we propose the following ansatz for constructing approximate Virasoro operators.

\begin{figure}[ht]
\centering
\includegraphics[width=\linewidth]{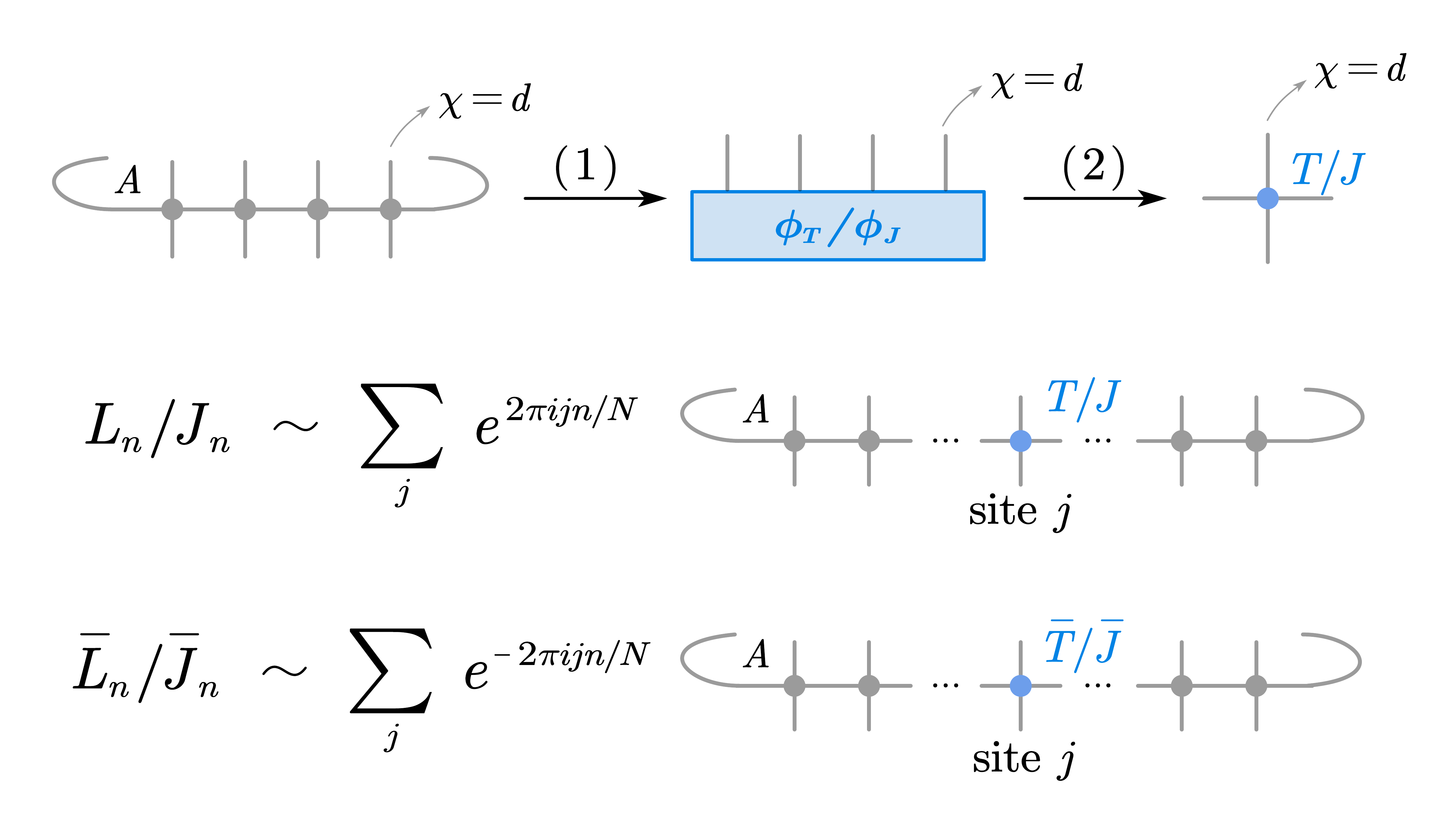}
\caption{Tensor network implementation of Eq.~\eqref{eq:Vir}. Eigenstate $\ket{\phi_T}$ or $\ket{\phi_J}$ that corresponds to energy-momentum tensor or Kac--Moody current is solved using a cylinder of 4 sites and reshaped into the four-leg blue tensor $T$ or $J$. The blue tensor $T$/$J$ is inserted at site $j$ and we sum over the contributions of different site $j$'s weighted by appropriate Fourier modes on the ring.}
\label{fig:Vir}
\end{figure}

Consider a generic tensor network made up of local tensor $A_{ijkl}$ of bond dimension $\chi_A=d$. We apply exact diagonalization on a small cylinder of four copies of tensor $A$ and identify the eigenstate $\ket{\phi_T}$ that corresponds to the energy-momentum tensor. To reach satisfactory accuracy of the spectrum with a cylinder of just 4 sites, coarse-graining schemes such as direct blocking, TRG or TNR may be required to obtain this local tensor $A$. This state $\ket{\phi_T}$ of size $d^{4}$ is then reshaped into a four-leg tensor of bond dimension $\chi = d$ so it could fit back into the lattice as the approximate $T$ operator, see Fig.~\ref{fig:Vir}.

Now using this approximate energy-momentum tensor $T$ obtained from cylinder of 4 sites, we can build the Virasoro generators for a larger cylinder of $N$ sites, as illustrated in Fig.~\ref{fig:Vir}. To represent $T(j)$, the energy-momentum tensor inserted at site $j$, we replace a local tensor $A$ at site $j$ in the transfer matrix with the approximate energy-momentum tensor $T$. The lattice Virasoro generator $L_n$ is then constructed as in Eq.~\eqref{eq:Vir}.

When the model possesses additional symmetries, the Virasoro symmetry in the underlying CFT theory can get extended to Kac--Moody symmetry. In those cases, the associated Kac--Moody generators on the lattice can be constructed as
\begin{equation}
\label{eq:KM}
J_n \sim \sum_{j=1}^N e^{\iu j n \frac{2\pi}{N}} J(j), \quad \bar{J}_n \sim \sum_{j=1}^N e^{-\iu j n \frac{2\pi}{N}} \bar{J}(j) .
\end{equation}
where $J$ and $\bar{J}$ are current operators with conformal dimensions $(1,0)$ and $(0,1)$.

Again we solve the eigenstate $\ket{\phi_J}$ that corresponds to the conserved current from a cylinder of 4 copies of $A$ by exact diagonlization. This vector $\ket{\phi_J}$ is subsequently reshaped into a four-leg operator and used to build the Kac--Moody generators $J_n$ on a larger cylinder, as described in Fig.~\ref{fig:Vir}.

\begin{figure}[ht]
\centering
\includegraphics[width=0.7\linewidth]{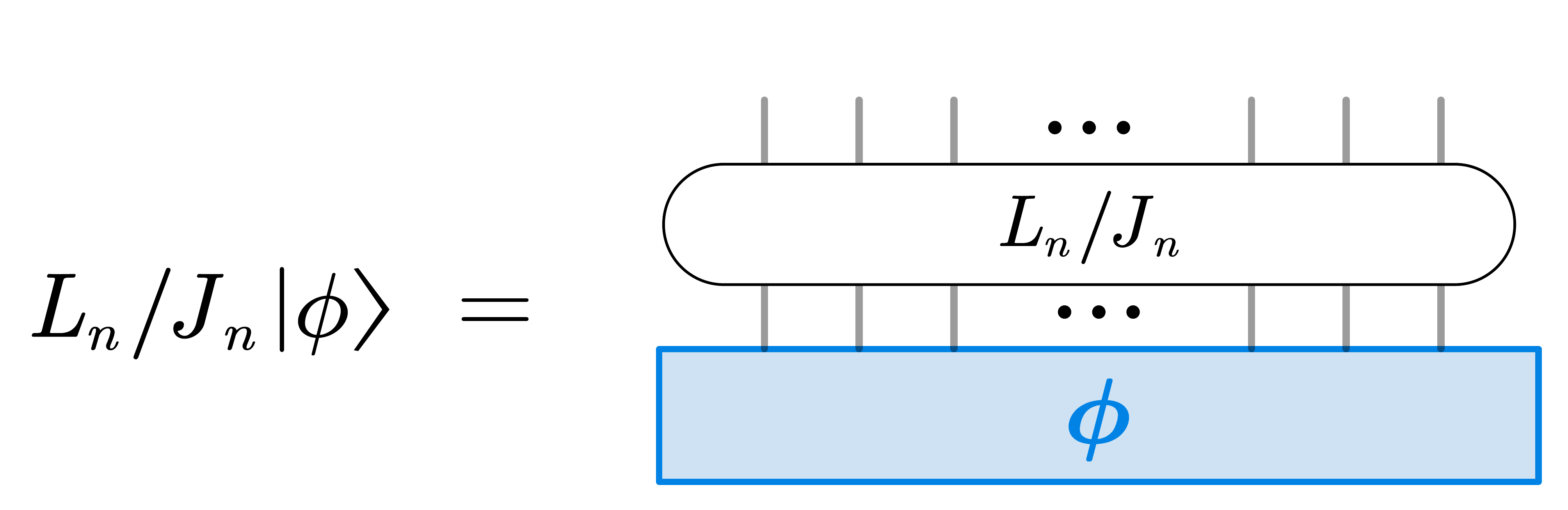}
\caption{Lattice Virasoro and Kac--Moody generators acting on eigenstates from the cylinder.}
\label{fig:operator}
\end{figure}

We check the validity of our proposal by applying the so constructed Virasoro/Kac--Moody generators to the low energy eigenstates of the transfer matrix of the Ising and the Dimer models, as to be explained in further details in the next section.

\section{Examples}

\subsection{Ising model \& Ising CFT}

As our first test, we study the Ising model on the square lattice $Z = \sum_{\langle ij \rangle} \exp{(-\beta \sigma_i \sigma_j)}$, where $\sigma \in \{-1, +1\}$ is an Ising spin. This partition function can be encoded into a square tensor network consisting of copies of local tensor $A^{(0)}_{ijkl}$ with components
\begin{equation}
\label{eq:isingtensor}
A^{(0)}_{ijkl} = \exp{ \left[ \beta (\sigma_i\sigma_j+\sigma_j\sigma_k+\sigma_k\sigma_l+\sigma_l\sigma_i) \right] }
\end{equation}
where $\beta=\beta_c=\log{(1+\sqrt{2})}/2$ at criticality.

\begin{figure}[ht]
\centering
\includegraphics[width=0.95\linewidth, left]{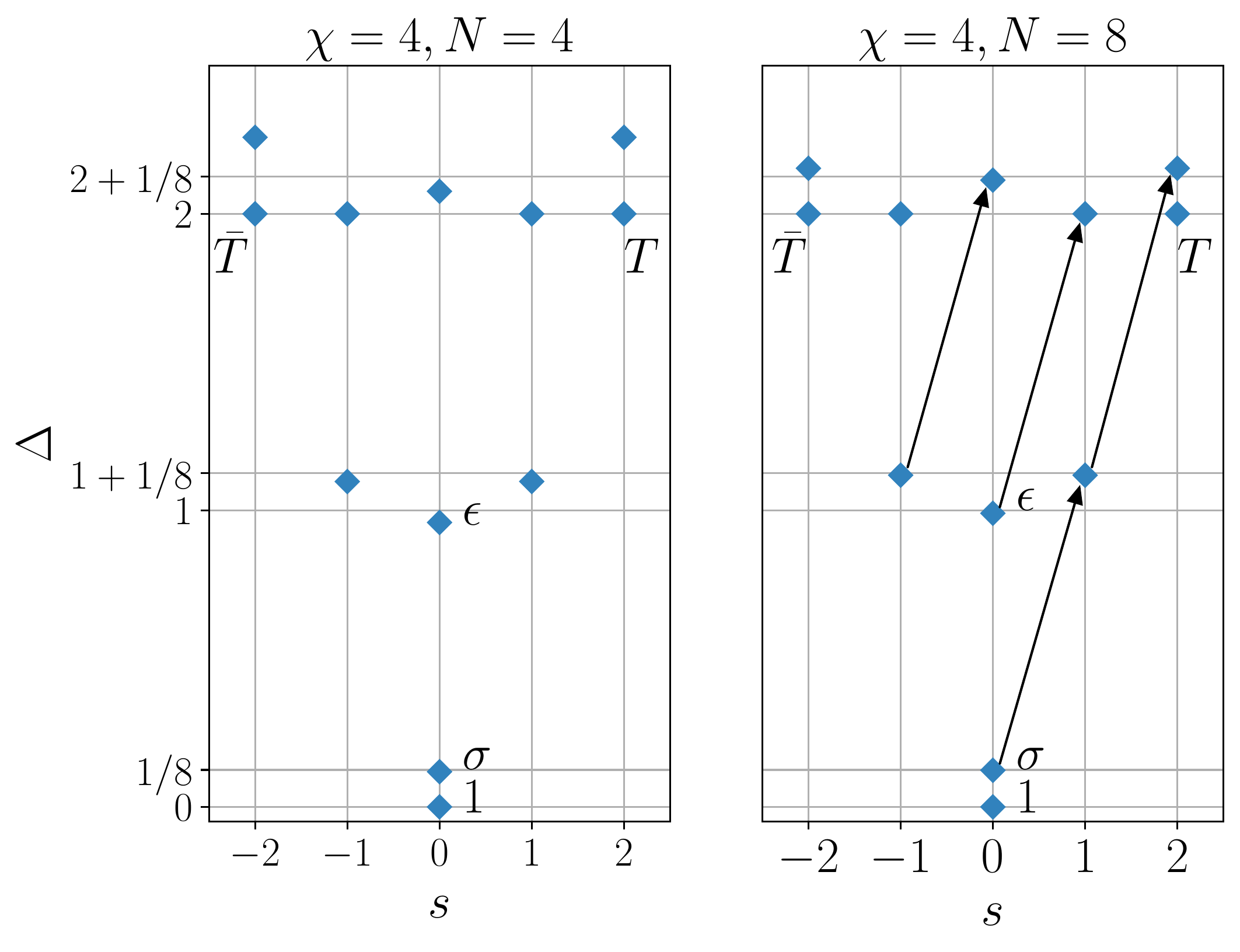}
\caption{Low energy spectra of Ising model in terms of scaling dimensions $\Delta$ and conformal spins $s$. Primary states are labeled with their corresponding primary operators in the CFT. We use the approximate lattice operators $T$ and $\bar{T}$ solved from a cylinder of 4 copies of $A$ (left) to construct lattice Virasoro generators $L_n$ and $\bar{L}_n$ for a cylinder of $N=8$ sites (right). As an example, actions of lattice $L_{-1}$ are illustrated on the right. Actions of other $L_n$ and $\bar{L}_{n}$ and the supporting matrix elements can be found in the appendices.}
\label{fig:A4spectrum}
\end{figure}

We start by blocking the tensor $A^{(0)}$ into $2\times2$ square block to form a new tensor $A$ of bond dimension $\chi=4$, which is used as the unit tensor in the following calculation. The idea is that we can use the eigenstates of small cylinder with $N=4$ copies of $A$ to construct the Virasoro generators $L_n$ and $\bar{L}_n$ for eigenstates of a larger cylinder (in this example, $N=8$). For both cylinders, low energy eigenstates together with conformal spins and scaling dimensions are obtained by exact diagonalization, see Fig.~\ref{fig:A4spectrum}.

In the spectrum of the small cylinder, we identify the state with conformal spin $s=2$ and scaling dimension $\Delta \approx 2$ as $\ket{\phi_T}$ that corresponds to the energy-momentum tensor $T$ and reshape it into a four-leg operator with bond dimension $\chi=4$. The approximate operator for the anti-holomorphic partner $\bar{T}$ is obtained similarly. (Technical subtlety involved in separating $\ket{\phi_T}$ and $\ket{\phi_{\bar{T}}}$ as degenerate states on a cylinder of 4 sites will be explained in the appendices.)

Using the approximate $T$ and $\bar{T}$ from cylinder of 4 sites, $L_n$ and $\bar{L}_n$ for the larger $N=8$ cylinder can be built via the construction as described in Fig.~\ref{fig:Vir} and Fig.~\ref{fig:operator}. The behaviors of $L_n$ and $L_{-n}$ are explored by examining the matrix elements $\mel{\phi_\beta}{L_n}{\phi_\alpha}$ in the low energy subspace of the $N=8$ transfer matrix.

In the CFT, Virasoro generators $L_n$ act as ladder operators on eigenstates, $L_n \ket{\Delta_\alpha, s_\alpha} \propto \ket{\Delta_\alpha - n, s_\alpha - n}$, raising $\Delta$ for $n<0$ and lowering $\Delta$ for $n>0$. Our numerical results show that, overall $L_n$ and $\bar{L}_{n}$ successfully map the states to their descendants as predicted by CFT with high fidelity $\frac{\lVert \mel{\phi_\beta}{L_n}{\phi_\alpha} \rVert}{\lVert \ket{\phi_\beta}\rVert \cdot \lVert {L_n}\ket{\phi_\alpha}\rVert} \gtrsim 0.9$ (tested with $n=\pm1, \pm2$ on cylinder of $N=8$ sites and bond dimension $\chi=4$). We do observe few erroneous matrix elements (one or two orders of magnitude less than the correct matrix elements) within the same Virasoro towers due to the finite size effect of lattice model and that $T$ and $\bar{T}$ are approximated from a very small cylinder.

\subsection{Dimer model \& free boson CFT}

The second example is the dimer model~\cite{kasteleyn1961, kasteleyn1963, temperley1961} which can be mapped to a bosonic theory via the height mapping~\cite{Allegra:2014mgc, henley1997, Ioffe:1989kd}. The partition function of dimer model can be represented using a square lattice consisting of a local tensor $B^{(0)}_{ijkl}$ which has non-zero components $B^{(0)}_{1111}=B^{(0)}_{2211}=B^{(0)}_{2121}=B^{(0)}_{1212}=B^{(0)}_{2222}=1$ and $B^{(0)}_{1122}=2$
\footnote{This tensor can be obtained by applying one step of TRG on the standard dimer tensor $\tilde{B}_{ijkl}$ which has non-zero components $\tilde{B}_{1112}= \tilde{B}_{1121}=\tilde{B}_{1211} = \tilde{B}_{2111} = 1$.}.
In the continuum limit, this model corresponds to $c=1$ free boson CFT with vertex operators of scaling dimensions $\Delta_{e,m} = e^2 + \frac{1}{4} m^2$ and conformal spin $s_{e,m} = em$, where $e$ and $m$ are the electromagnetic charges in the Coulomb gas description of the dimer model. In addition to Virasoro symmetry, the continuum limit of dimer model exhibits an extended symmetry, Kac--Moody symmetry.

Since the construction of Virasoro generators has already been tested in the previous case, here we will just focus on the applications of our ansatz in constructing the Kac--Moody generators. Again, we take a $2\times2$ block of tensors $B^{(0)}$ to form a new tensor $B$ with bond dimension $\chi = 4$. From the spectrum of a cylinder of 4 copies of $B$, we identify the lowest state with conformal spin $s=1$ as $\ket{\phi_J}$ and the lowest state with $s=-1$ as $\ket{\phi_{\bar{J}}}$. They are then reshaped into four-leg tensors as approximate lattice current operators $J$ and $\bar{J}$. We construct lattice Kac-Moody generators using approximate $J$ and $\bar{J}$ as in Fig.~\ref{fig:Vir} and check their behaviors as in Fig.~\ref{fig:operator} on low energy eigenstates from cylinder of $N=8$ copies of $B$.
\begin{figure}[ht]
\centering
\includegraphics[width=0.95\linewidth, left]{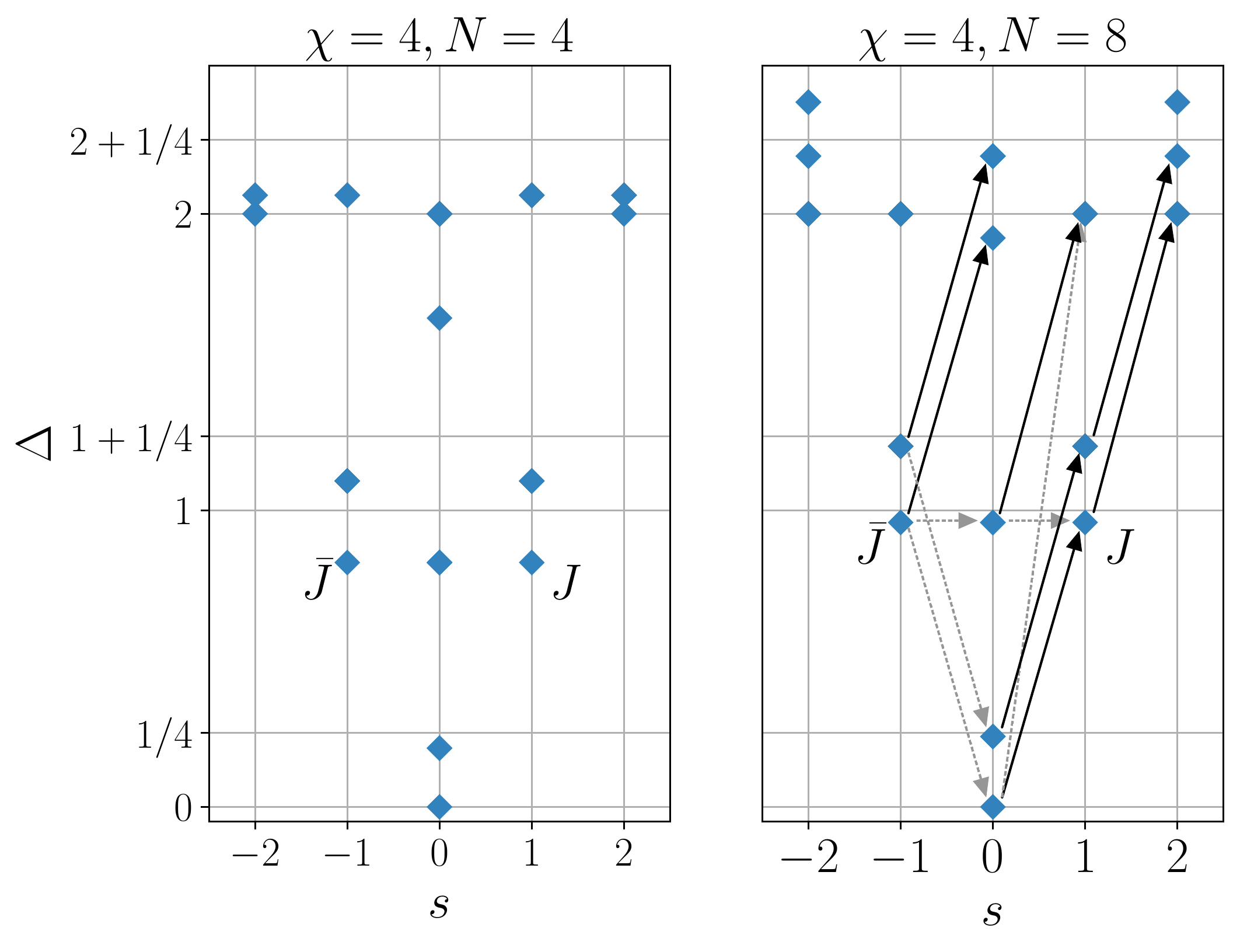}
\caption{Low energy spectra of dimer model in terms of scaling dimensions $\Delta$ and conformal spins $s$. Using approximate current operators $J$ and $\bar{J}$ obtained from a 4-site cylinder (left) we construct lattice Kac--Moody generators $J_n$ and $\bar{J}_n$ for a cylinder of $N=8$ sites (right) and examine their behaviors in the low energy subspace. Actions of lattice $J_{-1}$ that match with CFT predictions are illustrated using solid black arrows and the erroneous actions are marked with dashed gray arrows. See the appendices more examples.}
\label{fig:B4spectrum}
\end{figure}

In the CFT the Virasoro generators $L_n$ and Kac--Moody generators $J_n$ satisfy the commutation relation that $[L_m, J_n] = -n J_{m+n}$. Therefore $J_n$ act on eigenstates as ladder operators as well i.e. $J_n \ket{\Delta_\alpha, s_\alpha} \propto \ket{\Delta_\alpha - n, s_\alpha - n}$, raising $\Delta$ for $n<0$ and lowering $\Delta$ for $n>0$. We check the property of our lattice Kac--Moody generator by computing the matrix elements $\mel{\phi_\beta}{J_n}{\phi_\alpha}$. We confirm numerically that the correct Kac--Moody descendants $\ket{\phi_\beta} = \ket{\Delta_\beta, s_\beta}$ can be reached by acting $J_{n}$ and $\bar{J}_n$ on an initial state $\ket{\phi_\alpha} = \ket{\Delta_\alpha, s_\alpha}$. However, apart from those correct connections, errors in the matrix elements are also observed. In the case of dimer model, the errors are noticeably larger than those in the Ising case, as the dimer model suffers greater finite size effect. The errors fall in two classes: (i) the mixing of different Kac--Moody towers (mainly the tower of identity $\ket{s=0, \Delta=0}$ and that of $\ket{s=0, \Delta=1}$), which likely results from the fact that the lowest eigenstate of the cylinder is not the true vacuum but instead polluted by other primary states; (ii) the mixing of lowering and raising actions within the same Kac--Moody towers, suggesting that the holomorphic part of the lattice current and the anti-holomorphic part are not fully separated.
Having said that, one can refer to the appendix for the numerical results of the matrix element to see that the mixing is a consistent pattern of mixing -- meaning that the ground state is a definite mixing of the vacuum and several low lying degenerate excited states, and the mixing between $J$ and $\bar{J}$ is also consistent. Apart from this mixing, there is a 5-6 orders of magnitude separation in the matrix elements with states not supposed to be generated by the algebra from a given primary, suggesting that the method is working and that one can in principle isolate the eigenstates. 

To enhance the precision of this method, one could increase both the bond dimension $\chi$ of the local tensor in the small cylinder that is used for obtaining the lattice energy-momentum tensor or current, and lattice size $N$ of the cylinder on which the Virasoro/Kac--Moody generators are constructed. Tensor network techniques such as TNR can also be employed for approaching the fixed-point tensor of the model and therefore achieving a more accurate representation of the local energy-momentum tensor/current. It would also be interesting to combine the idea here with the scaling map approach as introduced in~\cite{Evenbly_2016}.

\section{Conclusion}

In this paper, we proposed a method to insert descendant operators in a tensor network representation of the partition function of a critical lattice model. The method was inspired by ``discrete holomorphicity'' --- observations made in 1980s that many CFT symmetries are actually preserved all the way up to the lattice scale (see~\cite{Cardy_2009} for a review). In this paper, we demonstrate that the stress tensor and the Kac--Moody current constructed from eigenstates of a tiny cylinder can be used to generate descendents of states defined on a big cylinder. In fact, in the examples considered, namely the Ising model and also the dimer model corresponding to a $c=1$ free boson CFT, we only need to use eigenstates of a cylinder with 4 sites to construct descendent states in a bigger cylinder. These conserved quantities seem to be defined at the smallest scale, mirroring the situation when these quantum operators were constructed to act on lattice states on a ring~\cite{Koo:1993wz, Milsted:2017csn, Zou:2017zce, Wang}. In the case of discrete holomorphic operators, their conservation follows from co-product structures of quantum algebra which are symmetries of the integrable lattice models involved. It would be very interesting to explore if a new algebraic structure is involved in the realization of the Virasoro and Kac--Moody symmetries.

Aside of its theoretical appeal, the method proposed here is, to our knowledge, the first method available to date to deal with the Virasoro and Kac--Moody symmetries when the quantum Hamiltonian is not readily accessible, an issue particularly pronounced in numerical computations. This should open the door to many applications.

\section{Acknowledgements}

LYH acknowledges the support of NSFC (Grant No. 11922502, 11875111) and the Shanghai Municipal Science and Technology Major Project
(Shanghai Grant No. 2019SHZDZX01), and Perimeter Institute for hospitality as a part of the Emmy Noether Fellowship programme.
We thank Lin Chen, Jiaqi Lou and Bingxin Lao for useful discussions and comments.

\bibliography{ref}

\begin{thebibliography}{22}%
\makeatletter
\providecommand \@ifxundefined [1]{%
 \@ifx{#1\undefined}
}%
\providecommand \@ifnum [1]{%
 \ifnum #1\expandafter \@firstoftwo
 \else \expandafter \@secondoftwo
 \fi
}%
\providecommand \@ifx [1]{%
 \ifx #1\expandafter \@firstoftwo
 \else \expandafter \@secondoftwo
 \fi
}%
\providecommand \natexlab [1]{#1}%
\providecommand \enquote  [1]{``#1''}%
\providecommand \bibnamefont  [1]{#1}%
\providecommand \bibfnamefont [1]{#1}%
\providecommand \citenamefont [1]{#1}%
\providecommand \href@noop [0]{\@secondoftwo}%
\providecommand \href [0]{\begingroup \@sanitize@url \@href}%
\providecommand \@href[1]{\@@startlink{#1}\@@href}%
\providecommand \@@href[1]{\endgroup#1\@@endlink}%
\providecommand \@sanitize@url [0]{\catcode `\\12\catcode `\$12\catcode
  `\&12\catcode `\#12\catcode `\^12\catcode `\_12\catcode `\%12\relax}%
\providecommand \@@startlink[1]{}%
\providecommand \@@endlink[0]{}%
\providecommand \url  [0]{\begingroup\@sanitize@url \@url }%
\providecommand \@url [1]{\endgroup\@href {#1}{\urlprefix }}%
\providecommand \urlprefix  [0]{URL }%
\providecommand \Eprint [0]{\href }%
\providecommand \doibase [0]{http://dx.doi.org/}%
\providecommand \selectlanguage [0]{\@gobble}%
\providecommand \bibinfo  [0]{\@secondoftwo}%
\providecommand \bibfield  [0]{\@secondoftwo}%
\providecommand \translation [1]{[#1]}%
\providecommand \BibitemOpen [0]{}%
\providecommand \bibitemStop [0]{}%
\providecommand \bibitemNoStop [0]{.\EOS\space}%
\providecommand \EOS [0]{\spacefactor3000\relax}%
\providecommand \BibitemShut  [1]{\csname bibitem#1\endcsname}%
\let\auto@bib@innerbib\@empty
\bibitem [{\citenamefont {Vanhove}\ \emph {et~al.}(2018)\citenamefont
  {Vanhove}, \citenamefont {Bal}, \citenamefont {Williamson}, \citenamefont
  {Bultinck}, \citenamefont {Haegeman},\ and\ \citenamefont
  {Verstraete}}]{Frank1}%
  \BibitemOpen
  \bibfield  {author} {\bibinfo {author} {\bibfnamefont {R.}~\bibnamefont
  {Vanhove}}, \bibinfo {author} {\bibfnamefont {M.}~\bibnamefont {Bal}},
  \bibinfo {author} {\bibfnamefont {D.~J.}\ \bibnamefont {Williamson}},
  \bibinfo {author} {\bibfnamefont {N.}~\bibnamefont {Bultinck}}, \bibinfo
  {author} {\bibfnamefont {J.}~\bibnamefont {Haegeman}}, \ and\ \bibinfo
  {author} {\bibfnamefont {F.}~\bibnamefont {Verstraete}},\ }\href {\doibase
  10.1103/PhysRevLett.121.177203} {\bibfield  {journal} {\bibinfo  {journal}
  {Phys. Rev. Lett.}\ }\textbf {\bibinfo {volume} {121}},\ \bibinfo {pages}
  {177203} (\bibinfo {year} {2018})}\BibitemShut {NoStop}%
\bibitem [{\citenamefont {Aasen}\ \emph {et~al.}(2020)\citenamefont {Aasen},
  \citenamefont {Fendley},\ and\ \citenamefont {Mong}}]{Aasen:2020jwb}%
  \BibitemOpen
  \bibfield  {author} {\bibinfo {author} {\bibfnamefont {D.}~\bibnamefont
  {Aasen}}, \bibinfo {author} {\bibfnamefont {P.}~\bibnamefont {Fendley}}, \
  and\ \bibinfo {author} {\bibfnamefont {R.~S.~K.}\ \bibnamefont {Mong}},\
  }\href@noop {} {\  (\bibinfo {year} {2020})},\ \Eprint
  {http://arxiv.org/abs/2008.08598} {arXiv:2008.08598 [cond-mat.stat-mech]}
  \BibitemShut {NoStop}%
\bibitem [{\citenamefont {Vanhove}\ \emph {et~al.}(2021)\citenamefont
  {Vanhove}, \citenamefont {Lootens}, \citenamefont {Van~Damme}, \citenamefont
  {Wolf}, \citenamefont {Osborne}, \citenamefont {Haegeman},\ and\
  \citenamefont {Verstraete}}]{frank_haagerup}%
  \BibitemOpen
  \bibfield  {author} {\bibinfo {author} {\bibfnamefont {R.}~\bibnamefont
  {Vanhove}}, \bibinfo {author} {\bibfnamefont {L.}~\bibnamefont {Lootens}},
  \bibinfo {author} {\bibfnamefont {M.}~\bibnamefont {Van~Damme}}, \bibinfo
  {author} {\bibfnamefont {R.}~\bibnamefont {Wolf}}, \bibinfo {author}
  {\bibfnamefont {T.}~\bibnamefont {Osborne}}, \bibinfo {author} {\bibfnamefont
  {J.}~\bibnamefont {Haegeman}}, \ and\ \bibinfo {author} {\bibfnamefont
  {F.}~\bibnamefont {Verstraete}},\ }\href {\doibase 10.48550/ARXIV.2110.03532}
  {\enquote {\bibinfo {title} {A critical lattice model for a haagerup
  conformal field theory},}\ } (\bibinfo {year} {2021})\BibitemShut {NoStop}%
\bibitem [{\citenamefont {Huang}\ \emph {et~al.}(2021)\citenamefont {Huang},
  \citenamefont {Lin}, \citenamefont {Ohmori}, \citenamefont {Tachikawa},\ and\
  \citenamefont {Tezuka}}]{Tachikawa_Haagerup}%
  \BibitemOpen
  \bibfield  {author} {\bibinfo {author} {\bibfnamefont {T.-C.}\ \bibnamefont
  {Huang}}, \bibinfo {author} {\bibfnamefont {Y.-H.}\ \bibnamefont {Lin}},
  \bibinfo {author} {\bibfnamefont {K.}~\bibnamefont {Ohmori}}, \bibinfo
  {author} {\bibfnamefont {Y.}~\bibnamefont {Tachikawa}}, \ and\ \bibinfo
  {author} {\bibfnamefont {M.}~\bibnamefont {Tezuka}},\ }\href {\doibase
  10.48550/ARXIV.2110.03008} {\enquote {\bibinfo {title} {Numerical evidence
  for a haagerup conformal field theory},}\ } (\bibinfo {year}
  {2021})\BibitemShut {NoStop}%
\bibitem [{\citenamefont {Levin}\ and\ \citenamefont
  {Nave}(2007)}]{Levin_2007}%
  \BibitemOpen
  \bibfield  {author} {\bibinfo {author} {\bibfnamefont {M.}~\bibnamefont
  {Levin}}\ and\ \bibinfo {author} {\bibfnamefont {C.~P.}\ \bibnamefont
  {Nave}},\ }\href {\doibase 10.1103/physrevlett.99.120601} {\bibfield
  {journal} {\bibinfo  {journal} {Physical Review Letters}\ }\textbf {\bibinfo
  {volume} {99}} (\bibinfo {year} {2007}),\
  10.1103/physrevlett.99.120601}\BibitemShut {NoStop}%
\bibitem [{\citenamefont {Evenbly}\ and\ \citenamefont
  {Vidal}(2015)}]{Evenbly_2015}%
  \BibitemOpen
  \bibfield  {author} {\bibinfo {author} {\bibfnamefont {G.}~\bibnamefont
  {Evenbly}}\ and\ \bibinfo {author} {\bibfnamefont {G.}~\bibnamefont
  {Vidal}},\ }\href {\doibase 10.1103/physrevlett.115.180405} {\bibfield
  {journal} {\bibinfo  {journal} {Physical Review Letters}\ }\textbf {\bibinfo
  {volume} {115}} (\bibinfo {year} {2015}),\
  10.1103/physrevlett.115.180405}\BibitemShut {NoStop}%
\bibitem [{\citenamefont {Koo}\ and\ \citenamefont
  {Saleur}(1994)}]{Koo:1993wz}%
  \BibitemOpen
  \bibfield  {author} {\bibinfo {author} {\bibfnamefont {W.~M.}\ \bibnamefont
  {Koo}}\ and\ \bibinfo {author} {\bibfnamefont {H.}~\bibnamefont {Saleur}},\
  }\href {\doibase 10.1016/0550-3213(94)90018-3} {\bibfield  {journal}
  {\bibinfo  {journal} {Nucl. Phys. B}\ }\textbf {\bibinfo {volume} {426}},\
  \bibinfo {pages} {459} (\bibinfo {year} {1994})},\ \Eprint
  {http://arxiv.org/abs/hep-th/9312156} {arXiv:hep-th/9312156} \BibitemShut
  {NoStop}%
\bibitem [{\citenamefont {Pasquier}\ and\ \citenamefont
  {Saleur}(1990)}]{Pasquier:1989kd}%
  \BibitemOpen
  \bibfield  {author} {\bibinfo {author} {\bibfnamefont {V.}~\bibnamefont
  {Pasquier}}\ and\ \bibinfo {author} {\bibfnamefont {H.}~\bibnamefont
  {Saleur}},\ }\href {\doibase 10.1016/0550-3213(90)90122-T} {\bibfield
  {journal} {\bibinfo  {journal} {Nucl. Phys. B}\ }\textbf {\bibinfo {volume}
  {330}},\ \bibinfo {pages} {523} (\bibinfo {year} {1990})}\BibitemShut
  {NoStop}%
\bibitem [{\citenamefont {Milsted}\ and\ \citenamefont
  {Vidal}(2017)}]{Milsted:2017csn}%
  \BibitemOpen
  \bibfield  {author} {\bibinfo {author} {\bibfnamefont {A.}~\bibnamefont
  {Milsted}}\ and\ \bibinfo {author} {\bibfnamefont {G.}~\bibnamefont
  {Vidal}},\ }\href {\doibase 10.1103/PhysRevB.96.245105} {\bibfield  {journal}
  {\bibinfo  {journal} {Phys. Rev. B}\ }\textbf {\bibinfo {volume} {96}},\
  \bibinfo {pages} {245105} (\bibinfo {year} {2017})},\ \Eprint
  {http://arxiv.org/abs/1706.01436} {arXiv:1706.01436 [cond-mat.str-el]}
  \BibitemShut {NoStop}%
\bibitem [{\citenamefont {Zou}\ \emph {et~al.}(2018)\citenamefont {Zou},
  \citenamefont {Milsted},\ and\ \citenamefont {Vidal}}]{Zou:2017zce}%
  \BibitemOpen
  \bibfield  {author} {\bibinfo {author} {\bibfnamefont {Y.}~\bibnamefont
  {Zou}}, \bibinfo {author} {\bibfnamefont {A.}~\bibnamefont {Milsted}}, \ and\
  \bibinfo {author} {\bibfnamefont {G.}~\bibnamefont {Vidal}},\ }\href
  {\doibase 10.1103/PhysRevLett.121.230402} {\bibfield  {journal} {\bibinfo
  {journal} {Phys. Rev. Lett.}\ }\textbf {\bibinfo {volume} {121}},\ \bibinfo
  {pages} {230402} (\bibinfo {year} {2018})},\ \Eprint
  {http://arxiv.org/abs/1710.05397} {arXiv:1710.05397 [cond-mat.str-el]}
  \BibitemShut {NoStop}%
\bibitem [{\citenamefont {Wang}\ \emph {et~al.}(2022)\citenamefont {Wang},
  \citenamefont {Zou},\ and\ \citenamefont {Vidal}}]{Wang}%
  \BibitemOpen
  \bibfield  {author} {\bibinfo {author} {\bibfnamefont {R.}~\bibnamefont
  {Wang}}, \bibinfo {author} {\bibfnamefont {Y.}~\bibnamefont {Zou}}, \ and\
  \bibinfo {author} {\bibfnamefont {G.}~\bibnamefont {Vidal}},\ }\href
  {\doibase 10.48550/ARXIV.2206.01656} {\enquote {\bibinfo {title} {Emergence
  of kac-moody symmetry in critical quantum spin chains},}\ } (\bibinfo {year}
  {2022})\BibitemShut {NoStop}%
\bibitem [{\citenamefont {Hauru}\ \emph {et~al.}(2016)\citenamefont {Hauru},
  \citenamefont {Evenbly}, \citenamefont {Ho}, \citenamefont {Gaiotto},\ and\
  \citenamefont {Vidal}}]{Hauru_2016}%
  \BibitemOpen
  \bibfield  {author} {\bibinfo {author} {\bibfnamefont {M.}~\bibnamefont
  {Hauru}}, \bibinfo {author} {\bibfnamefont {G.}~\bibnamefont {Evenbly}},
  \bibinfo {author} {\bibfnamefont {W.~W.}\ \bibnamefont {Ho}}, \bibinfo
  {author} {\bibfnamefont {D.}~\bibnamefont {Gaiotto}}, \ and\ \bibinfo
  {author} {\bibfnamefont {G.}~\bibnamefont {Vidal}},\ }\href {\doibase
  10.1103/PhysRevB.94.115125} {\bibfield  {journal} {\bibinfo  {journal} {Phys.
  Rev. B}\ }\textbf {\bibinfo {volume} {94}},\ \bibinfo {pages} {115125}
  (\bibinfo {year} {2016})}\BibitemShut {NoStop}%
\bibitem [{\citenamefont {Evenbly}\ and\ \citenamefont
  {Vidal}(2016)}]{Evenbly_2016}%
  \BibitemOpen
  \bibfield  {author} {\bibinfo {author} {\bibfnamefont {G.}~\bibnamefont
  {Evenbly}}\ and\ \bibinfo {author} {\bibfnamefont {G.}~\bibnamefont
  {Vidal}},\ }\href {\doibase 10.1103/physrevlett.116.040401} {\bibfield
  {journal} {\bibinfo  {journal} {Physical Review Letters}\ }\textbf {\bibinfo
  {volume} {116}} (\bibinfo {year} {2016}),\
  10.1103/physrevlett.116.040401}\BibitemShut {NoStop}%
\bibitem [{\citenamefont {Yang}\ \emph {et~al.}(2017)\citenamefont {Yang},
  \citenamefont {Gu},\ and\ \citenamefont {Wen}}]{Yang_2017}%
  \BibitemOpen
  \bibfield  {author} {\bibinfo {author} {\bibfnamefont {S.}~\bibnamefont
  {Yang}}, \bibinfo {author} {\bibfnamefont {Z.-C.}\ \bibnamefont {Gu}}, \ and\
  \bibinfo {author} {\bibfnamefont {X.-G.}\ \bibnamefont {Wen}},\ }\href
  {\doibase 10.1103/physrevlett.118.110504} {\bibfield  {journal} {\bibinfo
  {journal} {Physical Review Letters}\ }\textbf {\bibinfo {volume} {118}}
  (\bibinfo {year} {2017}),\ 10.1103/physrevlett.118.110504}\BibitemShut
  {NoStop}%
\bibitem [{\citenamefont {Kasteleyn}(1961)}]{kasteleyn1961}%
  \BibitemOpen
  \bibfield  {author} {\bibinfo {author} {\bibfnamefont {P.}~\bibnamefont
  {Kasteleyn}},\ }\href {\doibase https://doi.org/10.1016/0031-8914(61)90063-5}
  {\bibfield  {journal} {\bibinfo  {journal} {Physica}\ }\textbf {\bibinfo
  {volume} {27}},\ \bibinfo {pages} {1209} (\bibinfo {year}
  {1961})}\BibitemShut {NoStop}%
\bibitem [{\citenamefont {Kasteleyn}(1963)}]{kasteleyn1963}%
  \BibitemOpen
  \bibfield  {author} {\bibinfo {author} {\bibfnamefont {P.~W.}\ \bibnamefont
  {Kasteleyn}},\ }\href@noop {} {\bibfield  {journal} {\bibinfo  {journal}
  {Journal of Mathematical Physics}\ }\textbf {\bibinfo {volume} {4}},\
  \bibinfo {pages} {287} (\bibinfo {year} {1963})}\BibitemShut {NoStop}%
\bibitem [{\citenamefont {Temperley}\ and\ \citenamefont
  {Fisher}(1961)}]{temperley1961}%
  \BibitemOpen
  \bibfield  {author} {\bibinfo {author} {\bibfnamefont {H.~N.}\ \bibnamefont
  {Temperley}}\ and\ \bibinfo {author} {\bibfnamefont {M.~E.}\ \bibnamefont
  {Fisher}},\ }\href@noop {} {\bibfield  {journal} {\bibinfo  {journal}
  {Philosophical Magazine}\ }\textbf {\bibinfo {volume} {6}},\ \bibinfo {pages}
  {1061} (\bibinfo {year} {1961})}\BibitemShut {NoStop}%
\bibitem [{\citenamefont {Allegra}(2015)}]{Allegra:2014mgc}%
  \BibitemOpen
  \bibfield  {author} {\bibinfo {author} {\bibfnamefont {N.}~\bibnamefont
  {Allegra}},\ }\href {\doibase 10.1016/j.nuclphysb.2015.03.022} {\bibfield
  {journal} {\bibinfo  {journal} {Nucl. Phys. B}\ }\textbf {\bibinfo {volume}
  {894}},\ \bibinfo {pages} {685} (\bibinfo {year} {2015})},\ \Eprint
  {http://arxiv.org/abs/1410.4131} {arXiv:1410.4131 [cond-mat.stat-mech]}
  \BibitemShut {NoStop}%
\bibitem [{\citenamefont {Henley}(1997)}]{henley1997}%
  \BibitemOpen
  \bibfield  {author} {\bibinfo {author} {\bibfnamefont {C.~L.}\ \bibnamefont
  {Henley}},\ }\href {\doibase 10.1007/BF02765532} {\bibfield  {journal}
  {\bibinfo  {journal} {Journal of statistical physics}\ }\textbf {\bibinfo
  {volume} {89}},\ \bibinfo {pages} {483} (\bibinfo {year} {1997})},\ \Eprint
  {http://arxiv.org/abs/cond-mat/9607222} {arXiv:cond-mat/9607222 [cond-mat]}
  \BibitemShut {NoStop}%
\bibitem [{\citenamefont {Ioffe}\ and\ \citenamefont
  {Larkin}(1989)}]{Ioffe:1989kd}%
  \BibitemOpen
  \bibfield  {author} {\bibinfo {author} {\bibfnamefont {L.~B.}\ \bibnamefont
  {Ioffe}}\ and\ \bibinfo {author} {\bibfnamefont {A.~I.}\ \bibnamefont
  {Larkin}},\ }\href {\doibase 10.1103/PhysRevB.40.6941} {\bibfield  {journal}
  {\bibinfo  {journal} {Phys. Rev. B}\ }\textbf {\bibinfo {volume} {40}},\
  \bibinfo {pages} {6941} (\bibinfo {year} {1989})}\BibitemShut {NoStop}%
\bibitem [{Note1()}]{Note1}%
  \BibitemOpen
  \bibinfo {note} {This tensor can be obtained by applying one step of TRG on
  the standard dimer tensor $\protect \tilde {B}_{ijkl}$ which has non-zero
  components $\protect \tilde {B}_{1112}= \protect \tilde {B}_{1121}=\protect
  \tilde {B}_{1211} = \protect \tilde {B}_{2111} = 1$.}\BibitemShut {Stop}%
\bibitem [{\citenamefont {Cardy}(2009)}]{Cardy_2009}%
  \BibitemOpen
  \bibfield  {author} {\bibinfo {author} {\bibfnamefont {J.}~\bibnamefont
  {Cardy}},\ }\href {\doibase 10.1007/s10955-009-9870-6} {\bibfield  {journal}
  {\bibinfo  {journal} {Journal of Statistical Physics}\ }\textbf {\bibinfo
  {volume} {137}},\ \bibinfo {pages} {814} (\bibinfo {year}
  {2009})}\BibitemShut {NoStop}%
\end{thebibliography}%


%

\onecolumngrid

\appendix

\section{Appendices}

\subsection{Cylinder eigenstates and CFT spectra}

The partition function of a torus CFT can be written as:
\begin{align}
     Z_{\text{CFT}}
  &= \Tr \Bigl[
       \exp \bigl( -2\pi\Im\tau \cdot H_{\text{CFT}} \bigr)
       \exp \bigl(  2\pi\Re\tau \cdot P \bigr)
     \Bigr] \notag \\
  &= \Tr \Bigl[
       \exp \Bigl( -2\pi\Im\tau \Bigl( L_0 + \bar{L}_0 - \frac{c}{12} \Bigr) \Bigr)
       \exp \Bigl(  2\pi\Re\tau \bigl( L_0 - \bar{L}_0 \bigr) \Bigr)
     \Bigr].
\end{align}
Here, operators $H_{\text{CFT}}$ and $P$ denote the Hamiltonian and the total momentum operators that generate translations along the time and space directions, while $c$ is the central charge. $\tau$ is a complex modular parameter that parametrizes the geometry of the torus.

For a local field $\phi_\alpha$ with conformal dimension $(h_\alpha,\bar{h}_\alpha)$, from Virasogo algebra we know that
\begin{align}
     Z_{\text{CFT}}
  &= \sum_\alpha \exp \Bigl[
       - 2\pi    \Im\tau \Bigl( h_0 + \bar{h}_0 - \frac{c}{12} \Bigr)
       + 2\pi\iu \Re\tau \bigl( h_0 - \bar{h}_0 \bigr)
     \Bigr] \notag \\
  &= \sum_\alpha \exp \Bigl[
       - 2\pi    \Im\tau \Bigl(\Delta_\alpha - \frac{c}{12} \Bigr)
       + 2\pi\iu \Re\tau \cdot s_\alpha
     \Bigr],
\end{align}
where $\Delta_\alpha=h_0+\bar{h}_0$ and $s_\alpha=h_0-\bar{h}_0$ are the scaling dimension and conformal spin of $\phi_\alpha$ respectively.

In the continuum limit, a critical classical lattice model with periodic boundary conditions can be described by the torus CFT, with modular parameter $\tau=\iu m/N$. The partition function is then
\begin{equation}
  Z = \sum_\alpha \exp \Bigl[
      - 2\pi \frac{m}{N} \Bigl(\Delta_\alpha - \frac{c}{12} \Bigr)
      + mNf + \mathcal{O} \Bigl( \frac{m}{N^\gamma} \Bigr)
    \Bigr],
\end{equation}
where $f$ is the free energy per each site at the thermodynamic limit, and $\mathcal{O}(m/N^\gamma)$ is the finite-size corrections. If the partition function can be written as $Z=\Tr M^m$ where $M$ is the transfer matrix, the eigenvalues of $M$ are then
\begin{equation}
  \lambda_\alpha = \exp \Bigl[
    - \frac{2\pi}{N} \Bigl( \Delta_\alpha-\frac{c}{12} \Bigr)
    + Nf + \mathcal{O} \Bigl( \frac{1}{N^\gamma} \Bigr)
  \Bigr].
\end{equation}
The conformal spin can be calculated by introducing the translation operator $\exp(2\pi\iu P/N)$, whose eigenvalues are $\exp(2\pi\iu s_\alpha/N)$. For a translationally invariant model, $\exp(2\pi\iu P/N)$ commutes with the transfer matrix $M$, so they can be diagonalized simultaneously to produce a set of eigenvectors with eigenvalues $ \{ \lambda, P_\alpha \}$. One way of estimate the scaling dimension $\Delta_\alpha$ is to fix the identity state $\ket{\phi_I}$ and state $\ket{\phi_T}$ corresponding to energy momentum tensor to have $\Delta_I=0$ and $\Delta_T=2$ so that
\begin{equation}
\Delta_\alpha = \frac{2}{\log \lambda_T - \log \lambda_I} \left( \log \lambda_\alpha - \log \lambda_I \right) .
\end{equation}
Conformal spin is obtained from the eigenvalues of the lattice translation operator $P$ as
\begin{equation}
s_\alpha = \frac{N}{2\pi} P_\alpha .
\end{equation}

\subsection{Numerical details for the Ising model}

\definecolor{blue1}{RGB}{10,120,180}
\definecolor{blue2}{RGB}{107,175,215}
\newcommand{\good}{\color{blue1}\bfseries}
\newcommand{\bad}{\color{blue2}}

\begin{figure}[ht]
  \centering
  \includegraphics[width=0.5\textwidth]{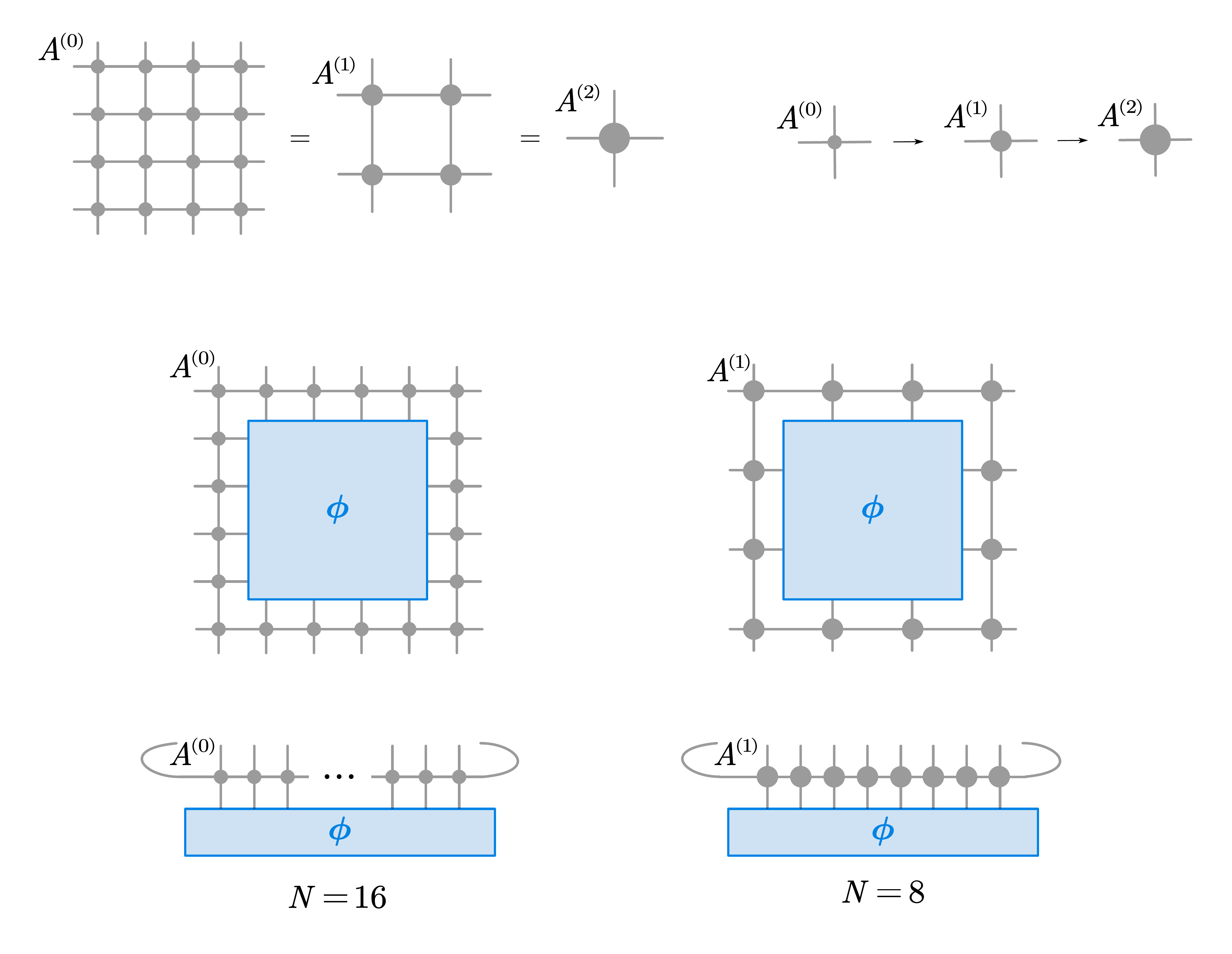}
  \caption{(Top) The $2\times2 \mapsto 1$ coarse-graining procedure. $A^{(s)}$ is made by blocking four copies of $A^{(s-1)}$. (Bottom) The corresponding operator insertion is effected by a square with fewer legs, each with a higher bond dimension.}
  \label{fig:rg}
\end{figure}

Here we provide the numerical details of the Ising model example discussed in the main text. In this example, we first apply one step of direct blocking as illustrated in Fig.~\ref{fig:rg} to obtain coarse-grained tensor $A=A^{(1)}$ of bond dimension $\chi = 4$ from standard square lattice Ising tensor $A^{(0)}$. More blocking steps or tensor network techniques such as TRG and TNR could be applied to obtain a unit tensor that is closer to the fixed-point, but here we find $A^{(1)}$ is sufficient for the purpose of illustrating our method. Low energy spectrum in terms of conformal spins and scaling dimensions can be obtained by exact diagonalization of transfer matrix on a cylinder, as discussed in last section.

Eigenstates $\ket{\phi_T}$ and $\ket{\phi_{\bar{T}}}$ that correspond to energy-momentum tensor $T$ and $\bar{T}$ have conformal spins $s=2$ and $s=-2$. Therefore on a cylinder of $N=4$ sites, these two eigenstates are in fact degenerate. In order to separate these two, one can go back to the original tensor $A^{(0)}$ of bond dimension $\chi=2$ and solve eigenstates from a cylinder of $N=8$ sites of $A^{(0)}$. These eigenstates are then reshaped into four-leg tensors $T$ and $\bar{T}$ with bond dimension $\chi=4$ and subsequently used to construct Virasoro generators $L_n$ and $L_{-n}$ on a cylinder of $N=8$ sites in the coarse-grained lattice of $A$.

\begin{figure}[htb]
  \centering
  \subfloat[$L_{-1}$]{
    \includegraphics[width=0.32\textwidth]{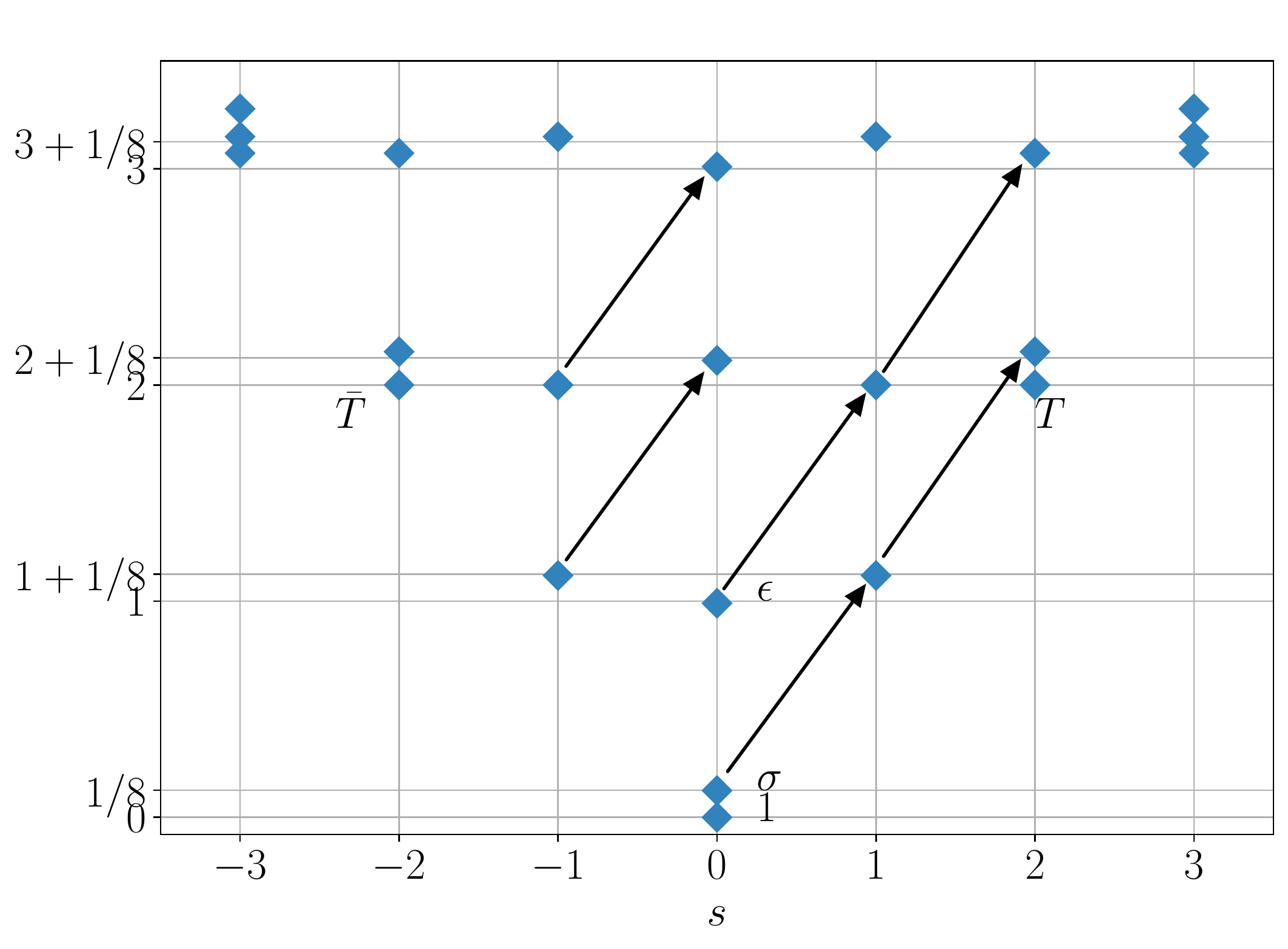}
    \label{fig:ising_lm1}}
  \qquad
  \subfloat[$\bar{L}_{-1}$]{
    \includegraphics[width=0.32\textwidth]{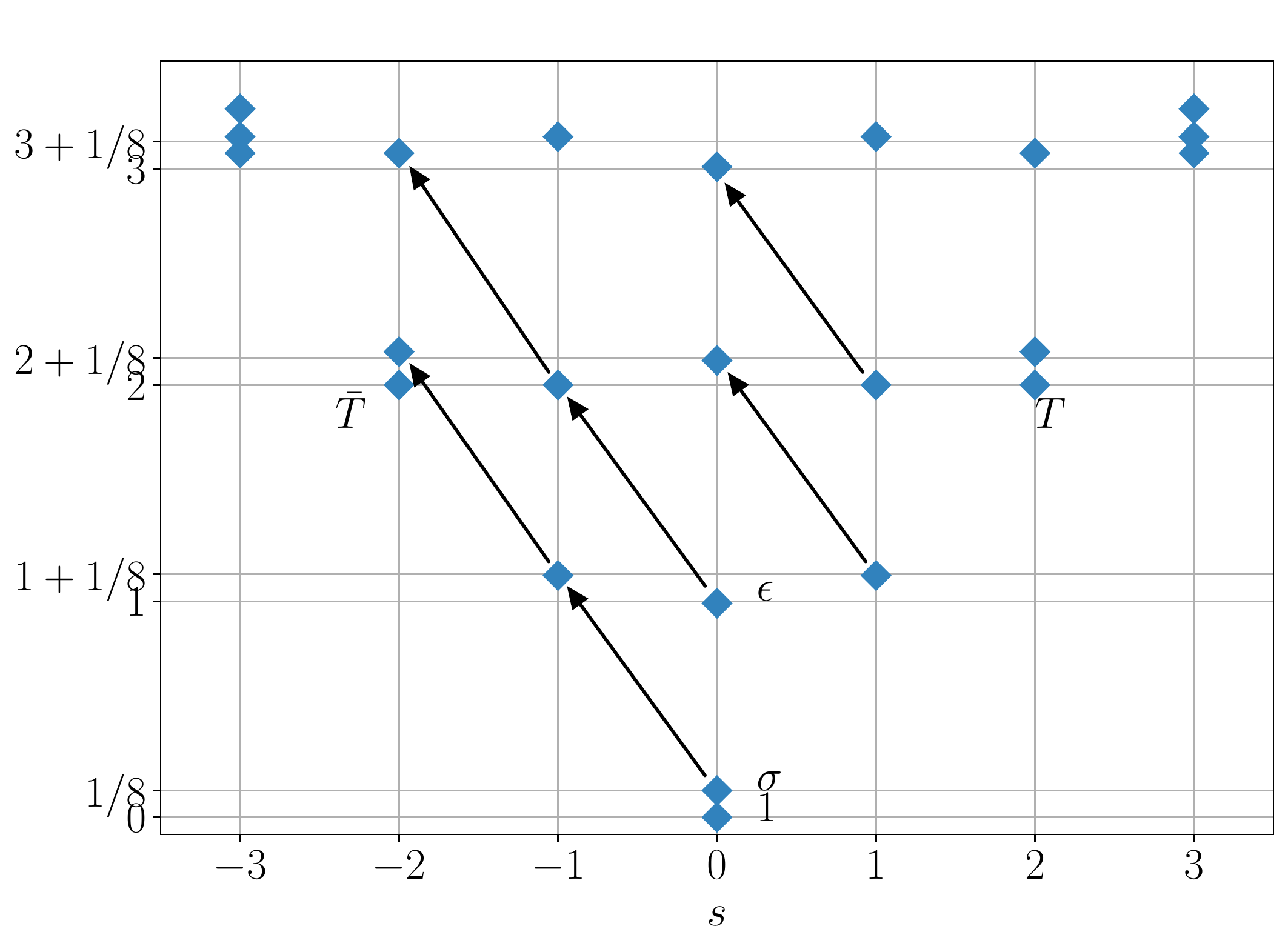}
    \label{fig:ising_lbarm1}}
  \\
  \subfloat[$L_{+1}$]{
    \includegraphics[width=0.32\textwidth]{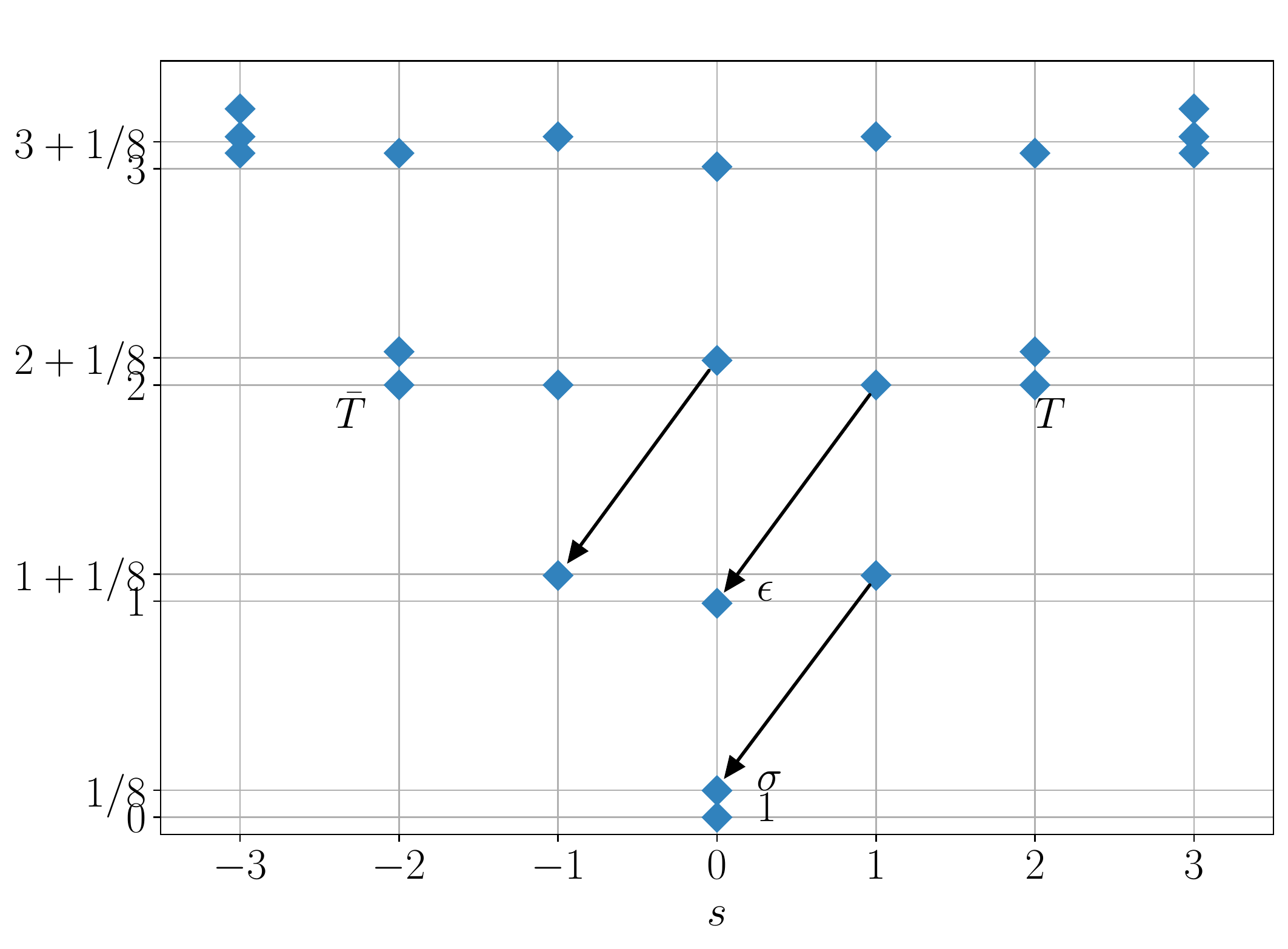}
    \label{fig:ising_l1}}
  \qquad
  \subfloat[$\bar{L}_{+1}$]{
    \includegraphics[width=0.32\textwidth]{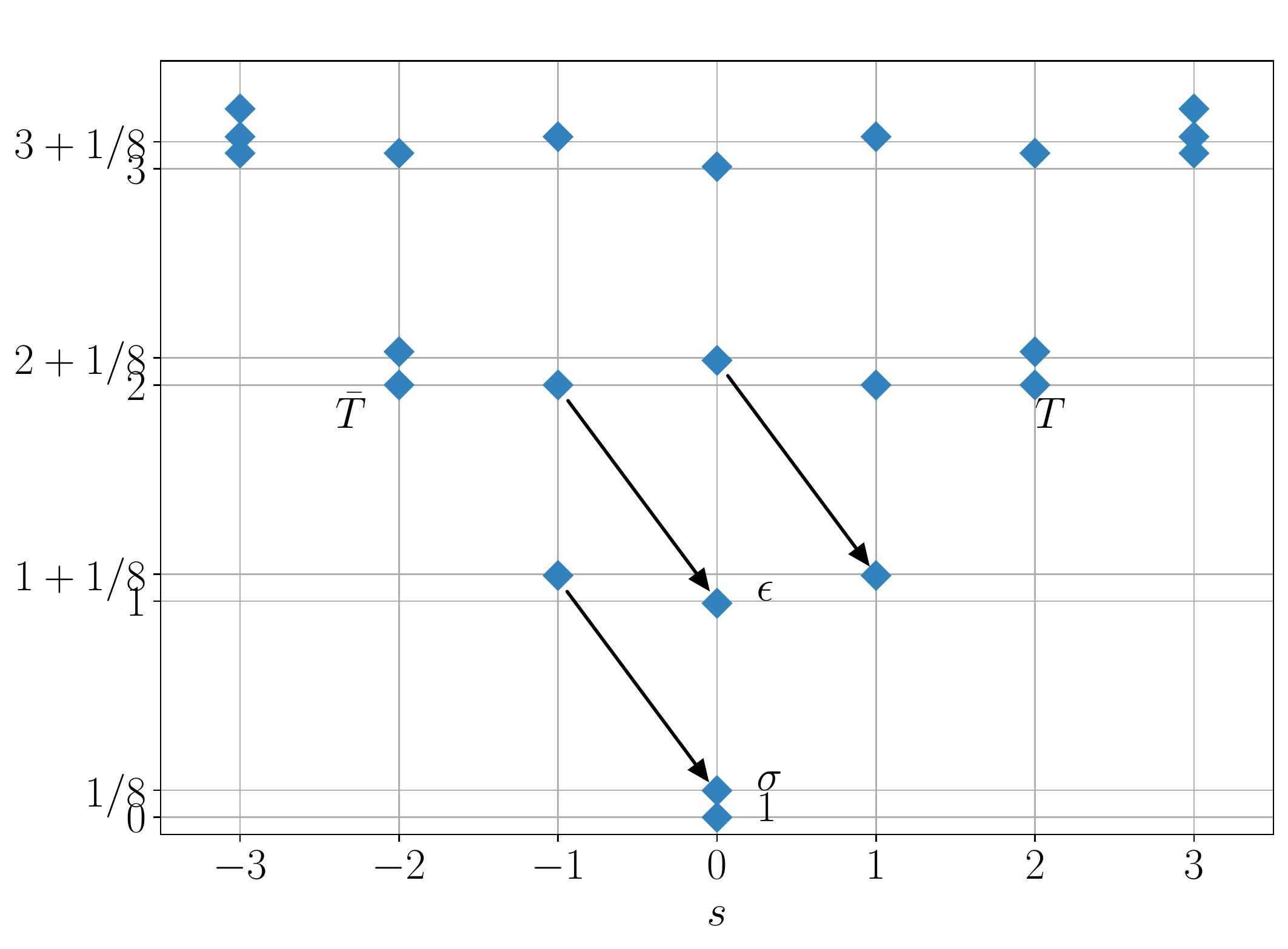}
    \label{fig:ising_lbar1}}
  \caption{Actions of lattice Virasoro generators $L_{\pm1}$ and $\bar{L}_{\pm1}$ in the low energy subspace of transfer matrix consisting of a row of $N=8$ copies of $A$ with bond dimension $\chi=4$. Corresponding matrix elements are listed in Table~\ref{tab:ising_lm1}--\ref{tab:ising_lbar1}. }
  \label{fig:l1}
\end{figure}

\begin{figure}[htb]
  \centering
  \subfloat[$L_{-2}$]{
    \includegraphics[width=0.32\textwidth]{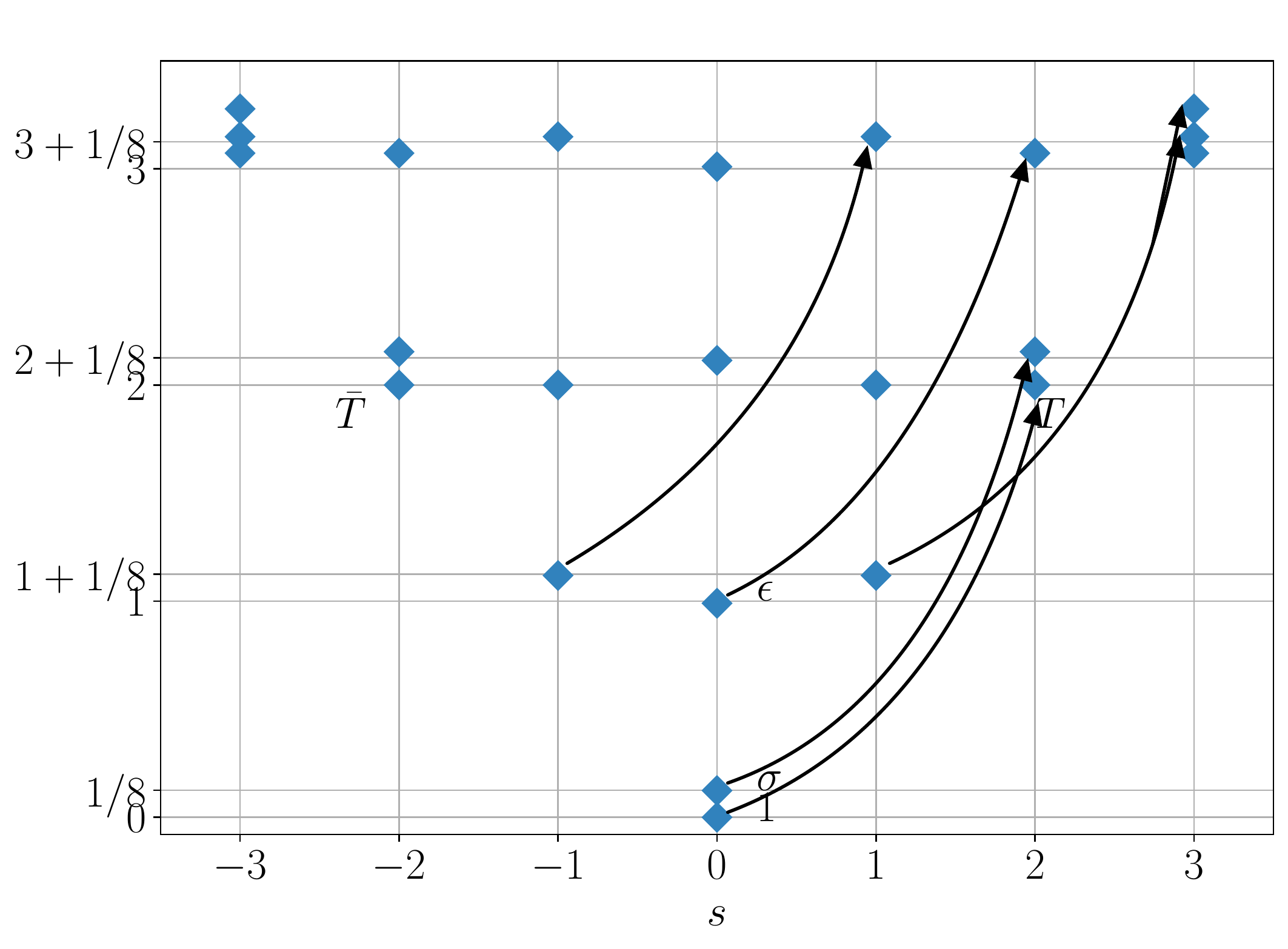}
    \label{fig:ising_lm2}}
  \qquad
  \subfloat[$\bar{L}_{-2}$]{
    \includegraphics[width=0.32\textwidth]{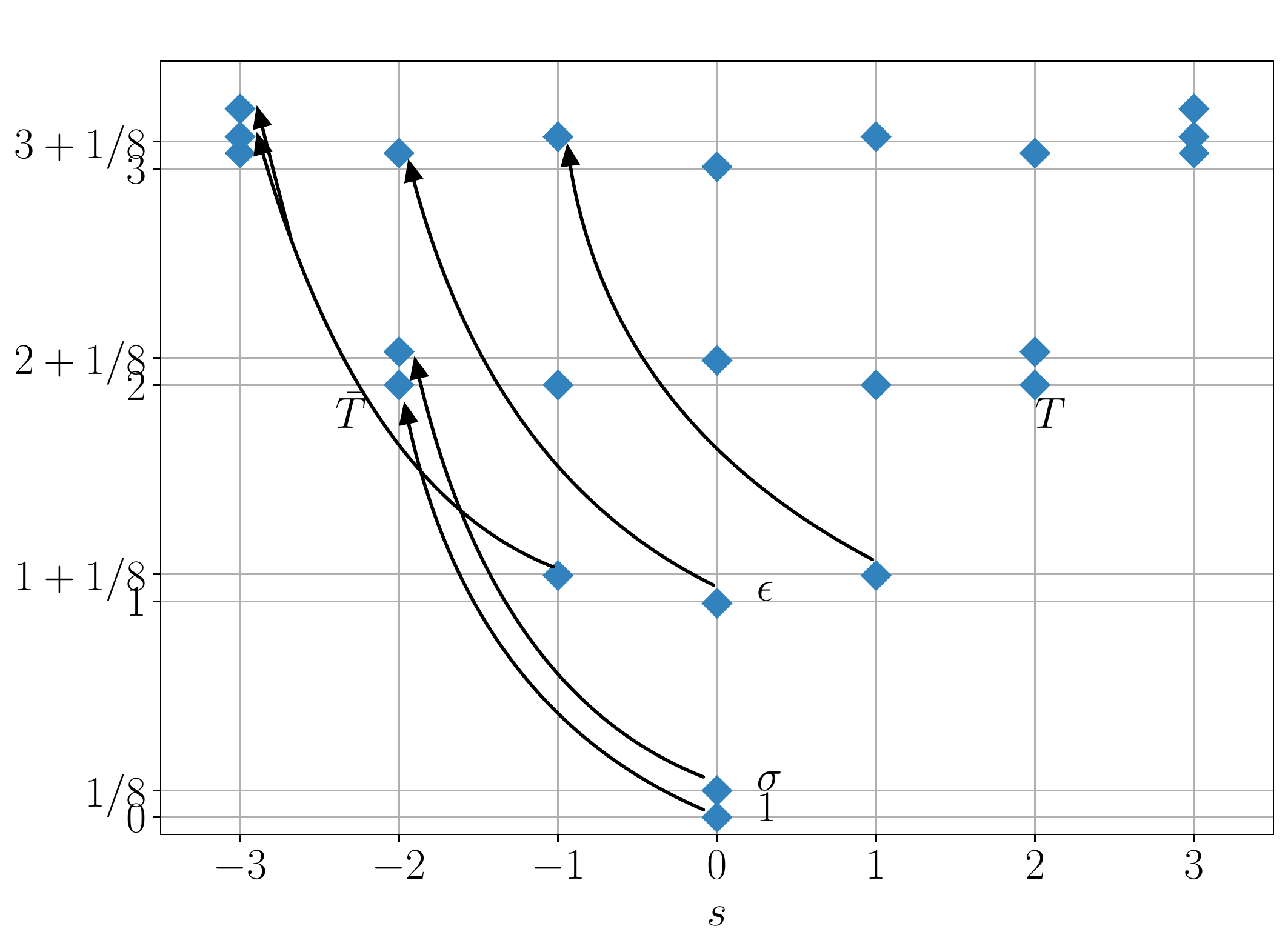}
    \label{fig:ising_lbarm2}}
  \\
  \subfloat[$L_{+2}$]{
    \includegraphics[width=0.32\textwidth]{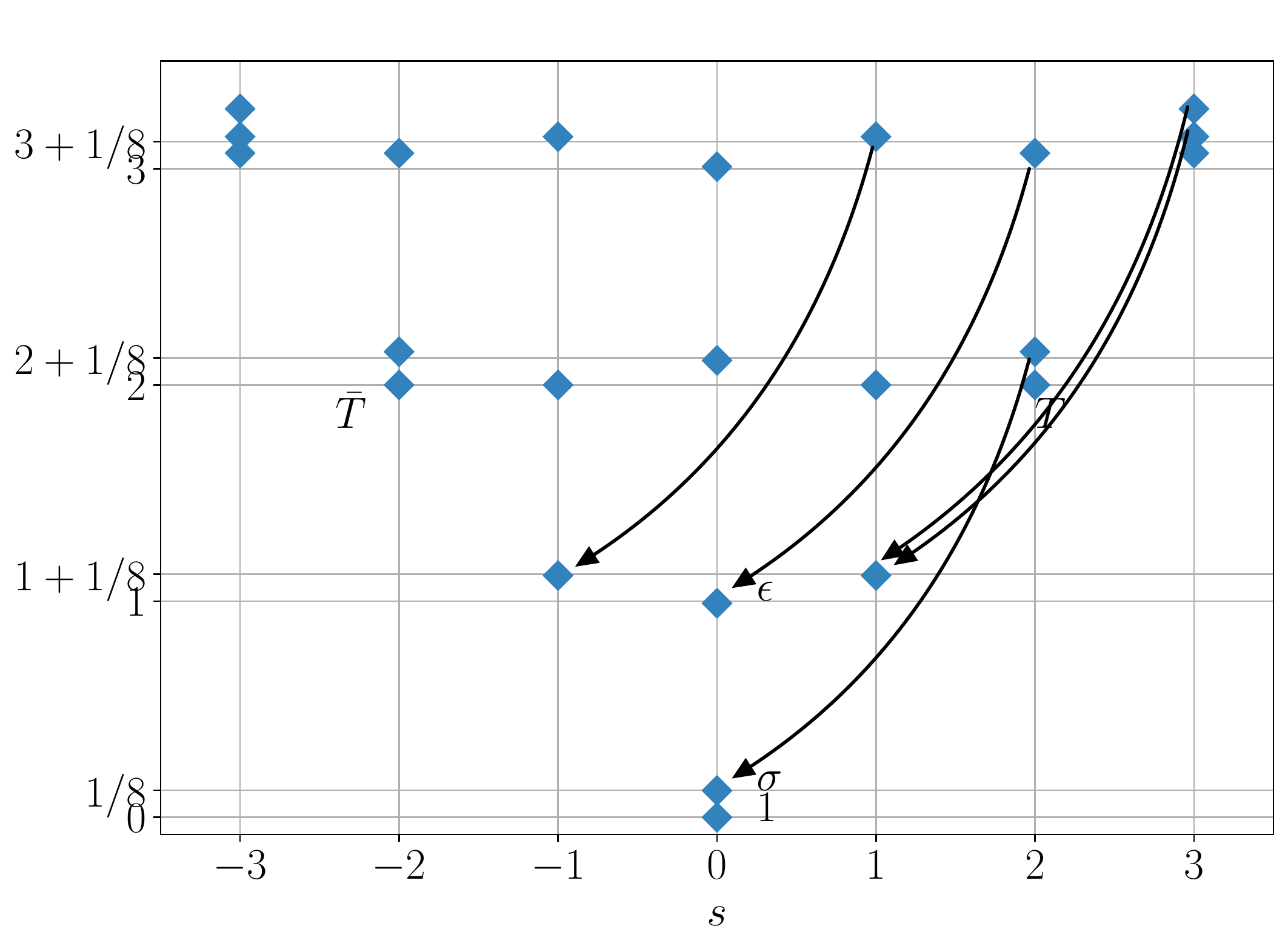}
    \label{fig:ising_l2}}
  \qquad
  \subfloat[$\bar{L}_{+2}$]{
    \includegraphics[width=0.32\textwidth]{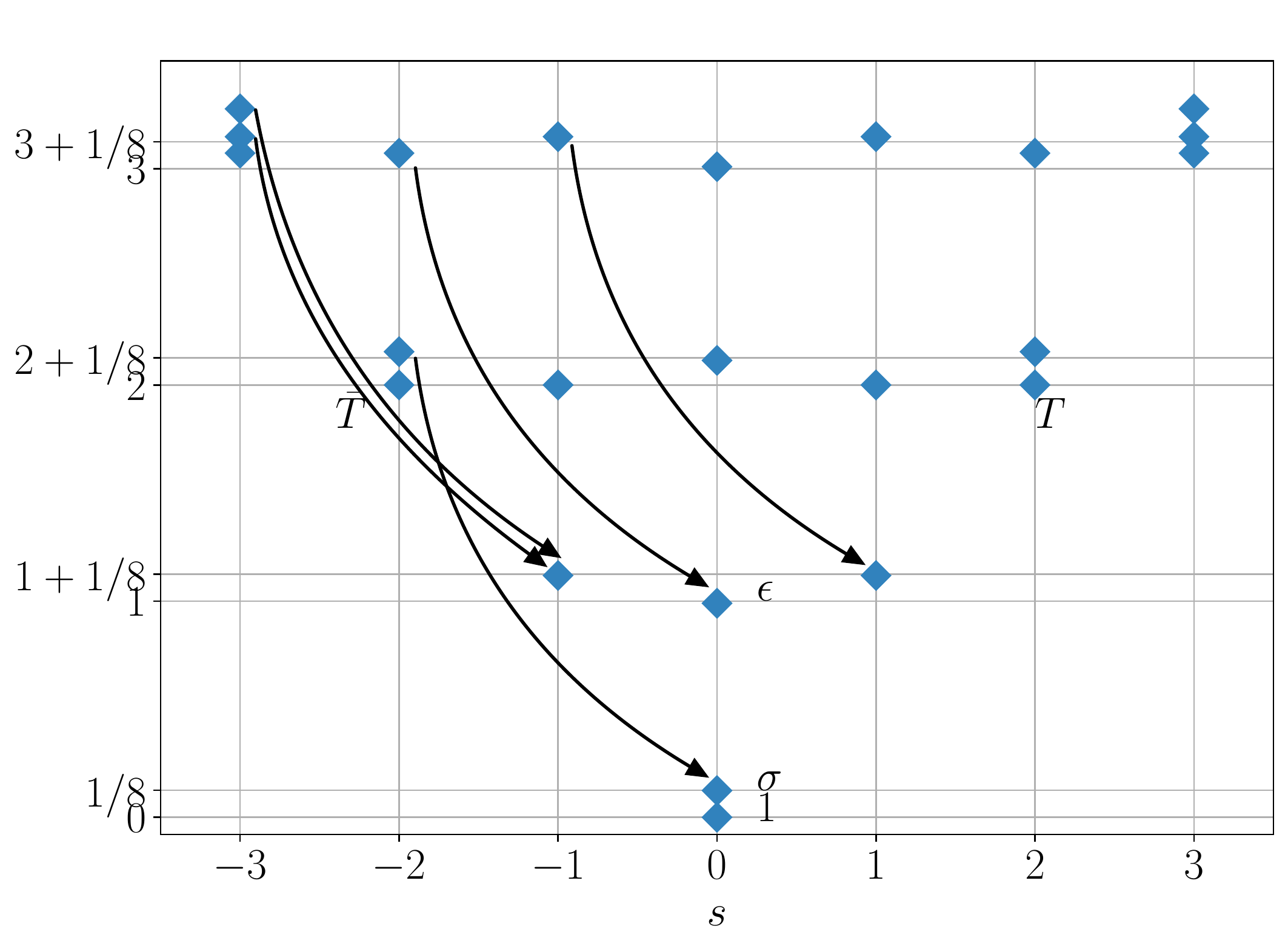}
    \label{fig:ising_lbar2}}
  \caption{Actions of lattice $L_{\pm2}$ and $\bar{L}_{\pm2}$ in the low energy subspace of transfer matrix consisting of a row of $N=8$ copies of $A$ with bond dimension $\chi=4$. Corresponding matrix elements are listed in Table~\ref{tab:ising_lm2}--\ref{tab:ising_lbar2}. }
  \label{fig:l2}
\end{figure}

We check the property of the lattice Virasoro generators by examining their actions on the low energy eigenstates of transfer matrix on the same cylinder of $N=8$ sites in the coarse-grained lattice. We list the matrix elements $\mel{\phi_\alpha}{L_n}{\phi_\beta}$ and $\mel{\phi_\alpha}{L_{-n}}{\phi_\beta}$ for $n=\pm1, \pm2$ in Table~\ref{tab:ising_lm1}--\ref{tab:ising_lbar2} and illustrate their behaviors on the conformal towers in Fig.~\ref{fig:l1} and~\ref{fig:l2}. In the tables, the matrix elements between states and their correct Virasoro descendant states are highlighted in bold blue, whereas the erroneous matrix elements are marked in pale blue.

\begin{table}[htb]
  \centering\small
  \def\M#1{$\mel{\phi_\alpha}{L_{-1}}{#1}$}
  \begin{tabular}{*{9}{c}}
    \toprule
    $\ket{\phi_\alpha}$ & $\Delta_\alpha$ & $(s_\alpha, \Delta^{\text{exact}}_\alpha)$ & \M2 & \M3 & \M4 & \M5 & \M6 & \M7 \\
    \midrule
    1  & 0.000 & (0, 0)      & 1.310e-16       & 4.976e-16       & 8.139e-17       & 7.343e-15       & 3.004e-15       & \bad 1.533e-01 \\
    2  & 0.123 & (0, 1/8)    & 5.482e-16       & 1.542e-16       & 7.609e-16       & \bad 8.919e-02  & 1.169e-17       & 9.638e-15 \\
    3  & 0.990 & (0, 1)      & 4.192e-15       & 5.818e-17       & 4.729e-17       & 9.432e-15       & 1.095e-16       & \bad 1.893e-03 \\
    4  & 1.119 & (1, 1+1/8)  & \good 8.968e-01 & 2.961e-15       & 4.735e-16       & 6.775e-16       & 4.944e-17       & 7.085e-17 \\
    5  & 1.119 & (-1, 1+1/8) & 1.112e-14       & 3.685e-16       & 6.305e-17       & 2.599e-17       & 2.711e-17       & 6.328e-17 \\
    6  & 2.000 & (1, 2)      & 3.966e-15       & \good 9.843e-01 & 5.230e-15       & 6.894e-15       & 3.497e-16       & 2.564e-16 \\
    7  & 2.000 & (-1, 2)     & 8.642e-17       & 4.463e-14       & 3.496e-15       & 3.084e-15       & 4.911e-16       & 3.656e-16 \\
    8  & 2.000 & (2, 2)      & 2.381e-17       & 3.123e-14       & 2.406e-15       & 1.607e-15       & \bad 3.717e-03  & 2.814e-16 \\
    9  & 2.000 & (-2, 2)     & 4.239e-17       & 3.613e-14       & 3.492e-15       & 1.603e-15       & 5.501e-16       & 4.390e-16 \\
    10 & 2.114 & (0, 2+1/8)  & 1.263e-16       & 1.149e-15       & 2.273e-14       & \good 9.136e-01 & 8.970e-16       & 2.874e-15 \\
    11 & 2.154 & (2, 2+1/8)  & 7.720e-16       & 9.948e-15       & \good 9.900e-01 & 3.329e-14       & 2.173e-15       & 1.253e-16 \\
    12 & 2.154 & (-2, 2+1/8) & 1.635e-16       & 3.915e-15       & 4.442e-14       & 1.977e-14       & 1.770e-15       & 2.668e-16 \\
    13 & 3.010 & (0, 3)      & 3.158e-17       & 2.695e-16       & 1.163e-16       & 3.422e-15       & 2.371e-14       & \good 9.723e-01 \\
    14 & 3.072 & (2, 3)      & 1.452e-16       & 2.926e-16       & 7.395e-16       & 1.135e-15       & \good 9.869e-01 & 9.407e-14 \\
    15 & 3.072 & (-2, 3)     & 1.580e-16       & 6.671e-17       & 8.238e-17       & 1.617e-16       & 3.911e-14       & 2.981e-16 \\
    16 & 3.072 & (3, 3)      & 3.441e-16       & 2.055e-16       & 4.705e-16       & 3.274e-16       & 4.410e-14       & 5.233e-14 \\
    17 & 3.072 & (-3, 3)     & 2.798e-16       & 3.365e-16       & 7.630e-17       & 9.554e-17       & 1.181e-14       & 2.835e-16 \\
    18 & 3.149 & (1, 3+1/8)  & \bad 2.156e-02  & 1.770e-15       & 3.732e-16       & 1.193e-16       & 2.812e-15       & 4.651e-15 \\
    19 & 3.149 & (-1, 3+1/8) & 1.457e-16       & 6.784e-17       & 4.529e-16       & 7.500e-17       & 6.743e-15       & 7.331e-16 \\
    20 & 3.149 & (3, 3+1/8)  & 5.871e-16       & 1.802e-16       & 1.424e-16       & 6.058e-16       & 4.553e-15       & 1.439e-15 \\
    21 & 3.149 & (-3, 3+1/8) & 1.086e-16       & 2.352e-16       & 2.470e-16       & 4.864e-16       & 2.529e-15       & 5.788e-16 \\
    22 & 3.277 & (3, 3+1/8)  & 5.969e-16       & 1.829e-16       & 3.332e-16       & 8.247e-16       & 4.104e-15       & 7.880e-15 \\
    23 & 3.277 & (-3, 3+1/8) & 3.814e-16       & 1.512e-16       & 1.246e-16       & 4.115e-16       & 2.339e-15       & 2.500e-16 \\
    24 & 4.000 & (0, 4)      & 3.229e-17       & 3.479e-16       & 1.744e-16       & 2.768e-15       & 7.413e-16       & \bad 3.735e-03 \\
    \bottomrule
  \end{tabular}
  \caption{\M{\phi_\beta}}
  \label{tab:ising_lm1}
\end{table}

\begin{table}[htb]
  \centering\small
  \def\M#1{$\mel{\phi_\alpha}{\bar{L}_{-1}}{#1}$}
  \begin{tabular}{*{9}{c}}
    \toprule
    $\ket{\phi_\alpha}$ & $\Delta_\alpha$ & $(s_\alpha, \Delta^{\text{exact}}_\alpha)$ & \M2 & \M3 & \M4 & \M5 & \M6 & \M7 \\
    \midrule
    1  & 0.000 & (0, 0)      & 1.310e-16        & 4.976e-16        & 7.343e-15        & 8.139e-17        & \bad 1.533e-01   & 3.004e-15 \\
    2  & 0.123 & (0, 1/8)    & 5.482e-16        & 1.542e-16        & \bad 8.919e-02   & 7.609e-16        & 9.638e-15        & 1.169e-17 \\
    3  & 0.990 & (0, 1)      & 4.192e-15        & 5.818e-17        & 9.432e-15        & 4.729e-17        & \bad 1.893e-03   & 1.095e-16 \\
    4  & 1.119 & (1, 1+1/8)  & 1.112e-14        & 3.685e-16        & 2.599e-17        & 6.305e-17        & 6.328e-17        & 2.711e-17 \\
    5  & 1.119 & (-1, 1+1/8) & \good  8.968e-01 & 2.961e-15        & 6.775e-16        & 4.735e-16        & 7.085e-17        & 4.944e-17 \\
    6  & 2.000 & (1, 2)      & 8.642e-17        & 4.463e-14        & 3.084e-15        & 3.496e-15        & 3.656e-16        & 4.911e-16 \\
    7  & 2.000 & (-1, 2)     & 3.966e-15        & \good  9.843e-01 & 6.894e-15        & 5.230e-15        & 2.564e-16        & 3.497e-16 \\
    8  & 2.000 & (2, 2)      & 4.239e-17        & 3.613e-14        & 1.603e-15        & 3.492e-15        & 4.390e-16        & 5.501e-16 \\
    9  & 2.000 & (-2, 2)     & 2.381e-17        & 3.123e-14        & 1.607e-15        & 2.406e-15        & 2.814e-16        & \bad 3.717e-03 \\
    10 & 2.114 & (0, 2+1/8)  & 1.263e-16        & 1.149e-15        & \good  9.136e-01 & 2.273e-14        & 2.874e-15        & 8.970e-16 \\
    11 & 2.154 & (2, 2+1/8)  & 1.635e-16        & 3.915e-15        & 1.977e-14        & 4.442e-14        & 2.668e-16        & 1.770e-15 \\
    12 & 2.154 & (-2, 2+1/8) & 7.720e-16        & 9.948e-15        & 3.329e-14        & \good  9.900e-01 & 1.253e-16        & 2.173e-15 \\
    13 & 3.010 & (0, 3)      & 3.158e-17        & 2.695e-16        & 3.422e-15        & 1.163e-16        & \good  9.723e-01 & 2.371e-14 \\
    14 & 3.072 & (2, 3)      & 1.580e-16        & 6.671e-17        & 1.617e-16        & 8.238e-17        & 2.981e-16        & 3.911e-14 \\
    15 & 3.072 & (-2, 3)     & 1.452e-16        & 2.926e-16        & 1.135e-15        & 7.395e-16        & 9.407e-14        & \good  9.869e-01 \\
    16 & 3.072 & (3, 3)      & 2.798e-16        & 3.365e-16        & 9.554e-17        & 7.630e-17        & 2.835e-16        & 1.181e-14 \\
    17 & 3.072 & (-3, 3)     & 3.441e-16        & 2.055e-16        & 3.274e-16        & 4.705e-16        & 5.233e-14        & 4.410e-14 \\
    18 & 3.149 & (1, 3+1/8)  & 1.457e-16        & 6.784e-17        & 7.500e-17        & 4.529e-16        & 7.331e-16        & 6.743e-15 \\
    19 & 3.149 & (-1, 3+1/8) & \bad 2.156e-02   & 1.770e-15        & 1.193e-16        & 3.732e-16        & 4.651e-15        & 2.812e-15 \\
    20 & 3.149 & (3, 3+1/8)  & 1.086e-16        & 2.352e-16        & 4.864e-16        & 2.470e-16        & 5.788e-16        & 2.529e-15 \\
    21 & 3.149 & (-3, 3+1/8) & 5.871e-16        & 1.802e-16        & 6.058e-16        & 1.424e-16        & 1.439e-15        & 4.553e-15 \\
    22 & 3.277 & (3, 3+1/8)  & 3.814e-16        & 1.512e-16        & 4.115e-16        & 1.246e-16        & 2.500e-16        & 2.339e-15 \\
    23 & 3.277 & (-3, 3+1/8) & 5.969e-16        & 1.829e-16        & 8.247e-16        & 3.332e-16        & 7.880e-15        & 4.104e-15 \\
    24 & 4.000 & (0, 4)      & 3.229e-17        & 3.479e-16        & 2.768e-15        & 1.744e-16        & \bad 3.735e-03   & 7.413e-16 \\
    \bottomrule
  \end{tabular}
  \caption{\M{\phi_\beta}}
  \label{tab:ising_lbarm1}
\end{table}

\begin{table}[htb]
  \centering\small
  \def\M#1{$\mel{\phi_\alpha}{L_{+1}}{#1}$}
  \makebox[0pt]{
    \begin{tabular}{*{12}{c}}
      \toprule
      $\ket{\phi_\alpha}$ & $\Delta_\alpha$ & $(s_\alpha, \Delta^{\text{exact}}_\alpha)$ & \M2 & \M3 & \M4 & \M5 & \M6 & \M7 & \M8 & \M9 & \M{10} \\
      \midrule
      1  & 0.000 & (0, 0)      & 8.578e-18      & 2.446e-16      & 2.604e-15       & 6.263e-18      & \bad 7.686e-03  & 3.004e-15      & 2.334e-15      & 1.933e-15      & 1.955e-16 \\
      2  & 0.123 & (0, 1/8)    & 1.552e-16      & 1.161e-14      & \good 9.555e-01 & 3.338e-14      & 4.367e-15       & 5.268e-16      & 1.785e-16      & 2.330e-16      & 1.878e-16 \\
      3  & 0.990 & (0, 1)      & 2.477e-16      & 2.052e-16      & 3.966e-15       & 1.351e-15      & \good 9.921e-01 & 3.729e-13      & 3.098e-13      & 2.601e-13      & 3.469e-15 \\
      4  & 1.119 & (1, 1+1/8)  & 5.814e-16      & 2.309e-16      & 7.517e-16       & 1.113e-15      & 7.151e-15       & 3.756e-14      & 3.285e-14      & 3.384e-14      & 9.396e-14 \\
      5  & 1.119 & (-1, 1+1/8) & \bad 4.909e-02 & 1.219e-14      & 8.994e-16       & 2.158e-16      & 2.381e-15       & 8.371e-15      & 5.368e-15      & 3.924e-15      & \good 9.454e-01 \\
      6  & 2.000 & (1, 2)      & 1.707e-17      & 7.965e-17      & 6.223e-17       & 9.033e-17      & 1.482e-16       & 1.990e-15      & \bad 3.293e-02 & 1.477e-15      & 2.462e-15 \\
      7  & 2.000 & (-1, 2)     & 4.645e-15      & \bad 3.095e-03 & 4.456e-17       & 1.054e-16      & 6.059e-17       & 5.274e-16      & 2.041e-15      & 7.986e-16      & 3.635e-15 \\
      8  & 2.000 & (2, 2)      & 7.978e-18      & 4.242e-16      & 5.599e-17       & 4.756e-17      & 7.192e-16       & 4.693e-16      & 9.823e-16      & 6.746e-16      & 9.167e-16 \\
      9  & 2.000 & (-2, 2)     & 1.314e-17      & 3.184e-16      & 4.905e-17       & 6.633e-15      & 8.610e-16       & \bad 3.055e-02 & 2.679e-15      & 6.948e-16      & 1.862e-17 \\
      10 & 2.114 & (0, 2+1/8)  & 2.846e-16      & 1.943e-16      & \bad 2.048e-02  & 7.711e-16      & 1.396e-15       & 7.353e-16      & 2.907e-16      & 2.730e-16      & 3.822e-16 \\
      11 & 2.154 & (2, 2+1/8)  & 1.532e-16      & 1.199e-16      & 3.187e-16       & 3.318e-16      & 2.437e-16       & 1.543e-15      & 1.025e-15      & 2.221e-16      & 4.611e-16 \\
      12 & 2.154 & (-2, 2+1/8) & 9.078e-17      & 9.965e-18      & 5.960e-16       & \bad 1.747e-03 & 7.664e-17       & 1.100e-14      & 4.836e-17      & 4.007e-17      & 4.441e-16 \\
      13 & 3.010 & (0, 3)      & 5.616e-18      & 5.907e-17      & 1.641e-15       & 7.723e-17      & \bad 1.303e-03  & 3.840e-16      & 5.947e-16      & 3.836e-16      & 1.136e-16 \\
      14 & 3.072 & (2, 3)      & 7.931e-16      & 4.752e-17      & 7.060e-17       & 6.990e-17      & 1.885e-16       & 4.479e-16      & 3.502e-16      & 2.643e-16      & 7.743e-17 \\
      15 & 3.072 & (-2, 3)     & 5.517e-16      & 1.185e-16      & 4.814e-17       & 2.421e-15      & 1.117e-16       & \bad 7.603e-03 & 3.901e-16      & 3.866e-16      & 5.368e-17 \\
      16 & 3.072 & (3, 3)      & 1.499e-15      & 3.170e-16      & 1.598e-17       & 2.964e-17      & 6.698e-16       & 3.871e-16      & 4.962e-16      & 3.717e-16      & 1.844e-17 \\
      17 & 3.072 & (-3, 3)     & 1.811e-15      & 2.346e-16      & 3.627e-17       & 4.429e-17      & 4.135e-16       & 2.970e-16      & 5.172e-16      & \bad 1.233e-02 & 4.658e-17 \\
      18 & 3.149 & (1, 3+1/8)  & 4.561e-16      & 9.261e-18      & 5.164e-16       & 6.611e-17      & 9.135e-18       & 2.588e-16      & 1.353e-15      & 2.115e-16      & 5.545e-16 \\
      19 & 3.149 & (-1, 3+1/8) & \bad 1.089e-01 & 6.010e-15      & 2.104e-16       & 6.268e-17      & 4.399e-17       & 2.843e-16      & 7.615e-17      & 1.238e-16      & \bad 2.463e-03 \\
      20 & 3.149 & (3, 3+1/8)  & 8.824e-16      & 1.183e-17      & 2.263e-17       & 3.453e-17      & 4.263e-17       & 1.052e-16      & 8.315e-17      & 1.835e-16      & 3.285e-16 \\
      21 & 3.149 & (-3, 3+1/8) & 2.440e-15      & 1.398e-17      & 4.274e-17       & 8.467e-17      & 3.126e-17       & 1.830e-16      & 2.468e-16      & 4.537e-15      & 6.787e-16 \\
      22 & 3.277 & (3, 3+1/8)  & 4.873e-16      & 6.136e-17      & 1.360e-16       & 1.070e-16      & 1.031e-16       & 6.228e-16      & 4.779e-16      & 3.234e-16      & 5.780e-16 \\
      23 & 3.277 & (-3, 3+1/8) & 1.049e-15      & 2.730e-17      & 2.764e-16       & 1.366e-16      & 3.310e-17       & 1.044e-16      & 1.319e-16      & 1.019e-15      & 3.057e-16 \\
      24 & 4.000 & (0, 4)      & 3.973e-17      & 1.041e-16      & 1.372e-15       & 1.821e-16      & \bad 1.508e-02  & 5.663e-15      & 4.731e-15      & 3.961e-15      & 1.292e-16 \\
      \bottomrule
    \end{tabular}
  }
  \caption{\M{\phi_\beta}}
  \label{tab:ising_l1}
\end{table}

\begin{table}[htb]
  \centering\small
  \def\M#1{$\mel{\phi_\alpha}{\bar{L}_{+1}}{#1}$}
  \makebox[0pt]{
    \begin{tabular}{*{12}{c}}
      \toprule
      $\ket{\phi_\alpha}$ & $\Delta_\alpha$ & $(s_\alpha, \Delta^{\text{exact}}_\alpha)$ & \M2 & \M3 & \M4 & \M5 & \M6 & \M7 & \M8 & \M9 & \M{10} \\
      \midrule
      1  & 0.000 & (0, 0)      & 8.578e-18      & 2.446e-16      & 6.263e-18      & 2.604e-15       & 3.004e-15      & \bad 7.686e-03  & 1.933e-15      & 2.334e-15      & 1.955e-16 \\
      2  & 0.123 & (0, 1/8)    & 1.552e-16      & 1.161e-14      & 3.338e-14      & \good 9.555e-01 & 5.268e-16      & 4.367e-15       & 2.330e-16      & 1.785e-16      & 1.878e-16 \\
      3  & 0.990 & (0, 1)      & 2.477e-16      & 2.052e-16      & 1.351e-15      & 3.966e-15       & 3.729e-13      & \good 9.921e-01 & 2.601e-13      & 3.098e-13      & 3.469e-15 \\
      4  & 1.119 & (1, 1+1/8)  & \bad 4.909e-02 & 1.219e-14      & 2.158e-16      & 8.994e-16       & 8.371e-15      & 2.381e-15       & 3.924e-15      & 5.368e-15      & \good 9.454e-01 \\
      5  & 1.119 & (-1, 1+1/8) & 5.814e-16      & 2.309e-16      & 1.113e-15      & 7.517e-16       & 3.756e-14      & 7.151e-15       & 3.384e-14      & 3.285e-14      & 9.396e-14 \\
      6  & 2.000 & (1, 2)      & 4.645e-15      & \bad 3.095e-03 & 1.054e-16      & 4.456e-17       & 5.274e-16      & 6.059e-17       & 7.986e-16      & 2.041e-15      & 3.635e-15 \\
      7  & 2.000 & (-1, 2)     & 1.707e-17      & 7.965e-17      & 9.033e-17      & 6.223e-17       & 1.990e-15      & 1.482e-16       & 1.477e-15      & \bad 3.293e-02 & 2.462e-15 \\
      8  & 2.000 & (2, 2)      & 1.314e-17      & 3.184e-16      & 6.633e-15      & 4.905e-17       & \bad 3.055e-02 & 8.610e-16       & 6.948e-16      & 2.679e-15      & 1.862e-17 \\
      9  & 2.000 & (-2, 2)     & 7.978e-18      & 4.242e-16      & 4.756e-17      & 5.599e-17       & 4.693e-16      & 7.192e-16       & 6.746e-16      & 9.823e-16      & 9.167e-16 \\
      10 & 2.114 & (0, 2+1/8)  & 2.846e-16      & 1.943e-16      & 7.711e-16      & \bad 2.048e-02  & 7.353e-16      & 1.396e-15       & 2.730e-16      & 2.907e-16      & 3.822e-16 \\
      11 & 2.154 & (2, 2+1/8)  & 9.078e-17      & 9.965e-18      & \bad 1.747e-03 & 5.960e-16       & 1.100e-14      & 7.664e-17       & 4.007e-17      & 4.836e-17      & 4.441e-16 \\
      12 & 2.154 & (-2, 2+1/8) & 1.532e-16      & 1.199e-16      & 3.318e-16      & 3.187e-16       & 1.543e-15      & 2.437e-16       & 2.221e-16      & 1.025e-15      & 4.611e-16 \\
      13 & 3.010 & (0, 3)      & 5.616e-18      & 5.907e-17      & 7.723e-17      & 1.641e-15       & 3.840e-16      & \bad 1.303e-03  & 3.836e-16      & 5.947e-16      & 1.136e-16 \\
      14 & 3.072 & (2, 3)      & 5.517e-16      & 1.185e-16      & 2.421e-15      & 4.814e-17       & \bad 7.603e-03 & 1.117e-16       & 3.866e-16      & 3.901e-16      & 5.368e-17 \\
      15 & 3.072 & (-2, 3)     & 7.931e-16      & 4.752e-17      & 6.990e-17      & 7.060e-17       & 4.479e-16      & 1.885e-16       & 2.643e-16      & 3.502e-16      & 7.743e-17 \\
      16 & 3.072 & (3, 3)      & 1.811e-15      & 2.346e-16      & 4.429e-17      & 3.627e-17       & 2.970e-16      & 4.135e-16       & \bad 1.233e-02 & 5.172e-16      & 4.658e-17 \\
      17 & 3.072 & (-3, 3)     & 1.499e-15      & 3.170e-16      & 2.964e-17      & 1.598e-17       & 3.871e-16      & 6.698e-16       & 3.717e-16      & 4.962e-16      & 1.844e-17 \\
      18 & 3.149 & (1, 3+1/8)  & \bad 1.089e-01 & 6.010e-15      & 6.268e-17      & 2.104e-16       & 2.843e-16      & 4.399e-17       & 1.238e-16      & 7.615e-17      & \bad 2.463e-03 \\
      19 & 3.149 & (-1, 3+1/8) & 4.561e-16      & 9.261e-18      & 6.611e-17      & 5.164e-16       & 2.588e-16      & 9.135e-18       & 2.115e-16      & 1.353e-15      & 5.545e-16 \\
      20 & 3.149 & (3, 3+1/8)  & 2.440e-15      & 1.398e-17      & 8.467e-17      & 4.274e-17       & 1.830e-16      & 3.126e-17       & 4.537e-15      & 2.468e-16      & 6.787e-16 \\
      21 & 3.149 & (-3, 3+1/8) & 8.824e-16      & 1.183e-17      & 3.453e-17      & 2.263e-17       & 1.052e-16      & 4.263e-17       & 1.835e-16      & 8.315e-17      & 3.285e-16 \\
      22 & 3.277 & (3, 3+1/8)  & 1.049e-15      & 2.730e-17      & 1.366e-16      & 2.764e-16       & 1.044e-16      & 3.310e-17       & 1.019e-15      & 1.319e-16      & 3.057e-16 \\
      23 & 3.277 & (-3, 3+1/8) & 4.873e-16      & 6.136e-17      & 1.070e-16      & 1.360e-16       & 6.228e-16      & 1.031e-16       & 3.234e-16      & 4.779e-16      & 5.780e-16 \\
      24 & 4.000 & (0, 4)      & 3.973e-17      & 1.041e-16      & 1.821e-16      & 1.372e-15       & 5.663e-15      & \bad 1.508e-02  & 3.961e-15      & 4.731e-15      & 1.292e-16 \\
      \bottomrule
    \end{tabular}
  }
  \caption{\M{\phi_\beta}}
  \label{tab:ising_lbar1}
\end{table}

\begin{table}[htb]
  \centering\small
  \def\M#1{$\mel{\phi_\alpha}{L_{-2}}{#1}$}
  \begin{tabular}{*{10}{c}}
    \toprule
    $\ket{\phi_\alpha}$ & $\Delta_\alpha$ & $(s_\alpha, \Delta^{\text{exact}}_\alpha)$ & \M1 & \M2 & \M3 & \M4 & \M5 & \M9 \\
    \midrule
    1  & 0.000 & (0, 0)      & 8.241e-17       & 6.521e-17       & 8.866e-17       & 2.950e-18       & 1.594e-17       & \bad 1.068e-03 \\
    2  & 0.123 & (0, 1/8)    & 6.633e-18       & 2.987e-16       & 6.894e-18       & 3.035e-16       & 7.635e-16       & 7.276e-15 \\
    3  & 0.990 & (0, 1)      & 1.788e-16       & 4.574e-17       & 3.614e-16       & 2.190e-17       & 2.721e-16       & \bad 3.095e-01 \\
    4  & 1.119 & (1, 1+1/8)  & 1.882e-15       & 2.846e-16       & 1.342e-16       & 7.095e-16       & \bad 7.017e-02  & 1.015e-16 \\
    5  & 1.119 & (-1, 1+1/8) & 1.536e-15       & 3.513e-16       & 2.972e-17       & 2.892e-16       & 9.902e-16       & 6.969e-17 \\
    6  & 2.000 & (1, 2)      & 6.049e-13       & 2.695e-15       & 2.314e-16       & 1.085e-16       & 8.067e-15       & 5.437e-15 \\
    7  & 2.000 & (-1, 2)     & 3.870e-13       & 3.393e-15       & 1.097e-17       & 1.124e-16       & 1.497e-16       & 6.186e-16 \\
    8  & 2.000 & (2, 2)      & \good 9.439e-01 & 2.002e-15       & \bad 5.724e-03  & 1.267e-16       & 8.287e-17       & 3.766e-16 \\
    9  & 2.000 & (-2, 2)     & 8.184e-13       & 3.389e-15       & 1.763e-16       & 1.227e-17       & 5.763e-17       & 1.272e-16 \\
    10 & 2.114 & (0, 2+1/8)  & 6.002e-16       & 1.925e-14       & 8.555e-16       & 1.343e-16       & 2.741e-16       & 2.290e-14 \\
    11 & 2.154 & (2, 2+1/8)  & 4.106e-15       & \good 9.608e-01 & 3.152e-15       & 7.502e-16       & 6.615e-16       & 3.385e-16 \\
    12 & 2.154 & (-2, 2+1/8) & 1.414e-15       & 4.282e-14       & 1.790e-15       & 3.127e-16       & 1.774e-16       & 5.431e-17 \\
    13 & 3.010 & (0, 3)      & 8.171e-16       & 1.306e-16       & 6.084e-16       & 5.863e-16       & 1.819e-16       & \bad 1.600e-02 \\
    14 & 3.072 & (2, 3)      & \bad 6.190e-03  & 3.048e-15       & \good 9.837e-01 & 2.157e-15       & 6.658e-15       & 1.768e-16 \\
    15 & 3.072 & (-2, 3)     & 1.383e-15       & 7.996e-17       & 4.043e-14       & 3.138e-15       & 7.003e-15       & 8.258e-16 \\
    16 & 3.072 & (3, 3)      & 1.339e-15       & 3.159e-17       & 2.026e-14       & 3.765e-15       & 1.605e-14       & 4.207e-15 \\
    17 & 3.072 & (-3, 3)     & 1.926e-16       & 7.461e-17       & 1.212e-14       & 2.186e-15       & 1.320e-14       & 3.003e-16 \\
    18 & 3.149 & (1, 3+1/8)  & 1.034e-16       & 3.715e-16       & 2.668e-15       & 9.258e-15       & \good 9.690e-01 & 6.707e-16 \\
    19 & 3.149 & (-1, 3+1/8) & 4.932e-17       & 2.229e-16       & 6.197e-15       & 1.526e-15       & 5.986e-15       & 9.006e-17 \\
    20 & 3.149 & (3, 3+1/8)  & 8.457e-17       & 2.925e-16       & 4.248e-15       & \bad 3.296e-01  & 3.429e-14       & 5.377e-16 \\
    21 & 3.149 & (-3, 3+1/8) & 1.105e-16       & 5.296e-17       & 2.522e-15       & 8.410e-15       & 7.983e-15       & 2.601e-16 \\
    22 & 3.277 & (3, 3+1/8)  & 1.928e-16       & 5.400e-16       & 5.464e-15       & \good 9.288e-01 & 3.894e-14       & 1.614e-15 \\
    23 & 3.277 & (-3, 3+1/8) & 7.712e-17       & 3.538e-16       & 2.324e-15       & 1.382e-14       & 4.187e-15       & 2.780e-16 \\
    24 & 4.000 & (0, 4)      & 1.765e-18       & 1.711e-16       & 7.672e-16       & 1.651e-16       & 6.216e-16       & \good 8.948e-01 \\
    \bottomrule
  \end{tabular}
  \caption{\M{\phi_\beta}}
  \label{tab:ising_lm2}
\end{table}

\begin{table}[htb]
  \centering\small
  \def\M#1{$\mel{\phi_\alpha}{\bar{L}_{-2}}{#1}$}
  \begin{tabular}{*{9}{c}}
    \toprule
    $\ket{\phi_\alpha}$ & $\Delta_\alpha$ & $(s_\alpha, \Delta^{\text{exact}}_\alpha)$ & \M1 & \M2 & \M3 & \M4 & \M5 & \M8 \\
    \midrule
    1  & 0.000 & (0, 0)      & 8.241e-17       & 6.521e-17       & 8.866e-17       & 1.594e-17       & 2.950e-18       & \bad 1.068e-03 \\
    2  & 0.123 & (0, 1/8)    & 6.633e-18       & 2.987e-16       & 6.894e-18       & 7.635e-16       & 3.035e-16       & 7.276e-15 \\
    3  & 0.990 & (0, 1)      & 1.788e-16       & 4.574e-17       & 3.614e-16       & 2.721e-16       & 2.190e-17       & \bad 3.095e-01 \\
    4  & 1.119 & (1, 1+1/8)  & 1.536e-15       & 3.513e-16       & 2.972e-17       & 9.902e-16       & 2.892e-16       & 6.969e-17 \\
    5  & 1.119 & (-1, 1+1/8) & 1.882e-15       & 2.846e-16       & 1.342e-16       & \bad 7.017e-02  & 7.095e-16       & 1.015e-16 \\
    6  & 2.000 & (1, 2)      & 3.870e-13       & 3.393e-15       & 1.097e-17       & 1.497e-16       & 1.124e-16       & 6.186e-16 \\
    7  & 2.000 & (-1, 2)     & 6.049e-13       & 2.695e-15       & 2.314e-16       & 8.067e-15       & 1.085e-16       & 5.437e-15 \\
    8  & 2.000 & (2, 2)      & 8.184e-13       & 3.389e-15       & 1.763e-16       & 5.763e-17       & 1.227e-17       & 1.272e-16 \\
    9  & 2.000 & (-2, 2)     & \good 9.439e-01 & 2.002e-15       & \bad 5.724e-03  & 8.287e-17       & 1.267e-16       & 3.766e-16 \\
    10 & 2.114 & (0, 2+1/8)  & 6.002e-16       & 1.925e-14       & 8.555e-16       & 2.741e-16       & 1.343e-16       & 2.290e-14 \\
    11 & 2.154 & (2, 2+1/8)  & 1.414e-15       & 4.282e-14       & 1.790e-15       & 1.774e-16       & 3.127e-16       & 5.431e-17 \\
    12 & 2.154 & (-2, 2+1/8) & 4.106e-15       & \good 9.608e-01 & 3.152e-15       & 6.615e-16       & 7.502e-16       & 3.385e-16 \\
    13 & 3.010 & (0, 3)      & 8.171e-16       & 1.306e-16       & 6.084e-16       & 1.819e-16       & 5.863e-16       & \bad 1.600e-02 \\
    14 & 3.072 & (2, 3)      & 1.383e-15       & 7.996e-17       & 4.043e-14       & 7.003e-15       & 3.138e-15       & 8.258e-16 \\
    15 & 3.072 & (-2, 3)     & \bad 6.190e-03  & 3.048e-15       & \good 9.837e-01 & 6.658e-15       & 2.157e-15       & 1.768e-16 \\
    16 & 3.072 & (3, 3)      & 1.926e-16       & 7.461e-17       & 1.212e-14       & 1.320e-14       & 2.186e-15       & 3.003e-16 \\
    17 & 3.072 & (-3, 3)     & 1.339e-15       & 3.159e-17       & 2.026e-14       & 1.605e-14       & 3.765e-15       & 4.207e-15 \\
    18 & 3.149 & (1, 3+1/8)  & 4.932e-17       & 2.229e-16       & 6.197e-15       & 5.986e-15       & 1.526e-15       & 9.006e-17 \\
    19 & 3.149 & (-1, 3+1/8) & 1.034e-16       & 3.715e-16       & 2.668e-15       & \good 9.690e-01 & 9.258e-15       & 6.707e-16 \\
    20 & 3.149 & (3, 3+1/8)  & 1.105e-16       & 5.296e-17       & 2.522e-15       & 7.983e-15       & 8.410e-15       & 2.601e-16 \\
    21 & 3.149 & (-3, 3+1/8) & 8.457e-17       & 2.925e-16       & 4.248e-15       & 3.429e-14       & \bad 3.296e-01  & 5.377e-16 \\
    22 & 3.277 & (3, 3+1/8)  & 7.712e-17       & 3.538e-16       & 2.324e-15       & 4.187e-15       & 1.382e-14       & 2.780e-16 \\
    23 & 3.277 & (-3, 3+1/8) & 1.928e-16       & 5.400e-16       & 5.464e-15       & 3.894e-14       & \good 9.288e-01 & 1.614e-15 \\
    24 & 4.000 & (0, 4)      & 1.765e-18       & 1.711e-16       & 7.672e-16       & 6.216e-16       & 1.651e-16       & \good 8.948e-01 \\
    \bottomrule
  \end{tabular}
  \caption{\M{\phi_\beta}}
  \label{tab:ising_lbarm2}
\end{table}

\begin{table}[htb]
  \centering\small
  \def\M#1{$\mel{\phi_\alpha}{L_{+2}}{#1}$}
  \begin{tabular}{*{9}{c}}
    \toprule
    $\ket{\phi_\alpha}$ & $\Delta_\alpha$ & $(s_\alpha, \Delta^{\text{exact}}_\alpha)$ & \M{11} & \M{14} & \M{18} & \M{20} & \M{22} & \M{23} \\
    \midrule
    1  & 0.000 & (0, 0)      & 5.467e-15       & \bad 4.795e-03  & 1.856e-16       & 1.520e-16       & 1.245e-16       & 5.713e-16 \\
    2  & 0.123 & (0, 1/8)    & \good 9.867e-01 & 4.046e-15       & 2.586e-16       & 5.058e-16       & 8.715e-16       & 9.216e-16 \\
    3  & 0.990 & (0, 1)      & 3.940e-15       & \good 9.955e-01 & 1.994e-14       & 1.384e-14       & 1.349e-15       & 1.628e-14 \\
    4  & 1.119 & (1, 1+1/8)  & 6.964e-16       & 2.776e-15       & 3.919e-13       & \good 9.842e-01 & \good 9.962e-01 & 2.353e-12 \\
    5  & 1.119 & (-1, 1+1/8) & 6.594e-16       & 2.919e-15       & \good 9.841e-01 & 4.570e-13       & 4.304e-13       & 1.000e-14 \\
    6  & 2.000 & (1, 2)      & 6.025e-17       & 3.039e-16       & 1.670e-16       & 4.959e-15       & 1.318e-15       & 5.919e-16 \\
    7  & 2.000 & (-1, 2)     & 2.966e-17       & 2.019e-16       & 3.288e-15       & 5.373e-16       & 1.743e-16       & 2.381e-15 \\
    8  & 2.000 & (2, 2)      & 7.495e-17       & 6.037e-16       & 3.122e-16       & 3.257e-16       & 3.502e-16       & 1.903e-15 \\
    9  & 2.000 & (-2, 2)     & 7.485e-17       & 2.251e-16       & 2.593e-16       & 2.083e-16       & 2.246e-16       & 4.436e-16 \\
    10 & 2.114 & (0, 2+1/8)  & \bad 8.935e-04  & 9.438e-16       & 3.131e-16       & 3.338e-16       & 2.725e-16       & 7.598e-16 \\
    11 & 2.154 & (2, 2+1/8)  & 7.313e-16       & 1.787e-16       & 3.496e-16       & 4.773e-16       & 4.931e-16       & 2.210e-15 \\
    12 & 2.154 & (-2, 2+1/8) & 6.331e-16       & 9.592e-17       & 2.916e-16       & 1.403e-16       & 2.540e-16       & 4.221e-16 \\
    13 & 3.010 & (0, 3)      & 1.460e-15       & 1.158e-03       & 6.726e-17       & 5.694e-17       & 8.429e-17       & 1.475e-15 \\
    14 & 3.072 & (2, 3)      & 4.996e-17       & 1.423e-16       & 2.763e-17       & 1.681e-16       & 4.588e-17       & 1.346e-15 \\
    15 & 3.072 & (-2, 3)     & 3.628e-17       & 2.232e-16       & 2.778e-17       & 1.042e-16       & 2.724e-17       & 1.397e-15 \\
    16 & 3.072 & (3, 3)      & 3.601e-17       & 3.837e-16       & 1.870e-16       & 4.195e-16       & 9.056e-17       & 1.606e-14 \\
    17 & 3.072 & (-3, 3)     & 3.907e-17       & 5.139e-16       & 5.574e-17       & 1.623e-16       & 1.383e-16       & 1.521e-15 \\
    18 & 3.149 & (1, 3+1/8)  & 5.707e-16       & 2.357e-17       & 1.371e-15       & \bad 1.438e-02  & \bad 2.638e-04  & 5.981e-15 \\
    19 & 3.149 & (-1, 3+1/8) & 3.890e-16       & 3.388e-17       & \bad 2.615e-03  & 2.466e-15       & 2.766e-17       & 2.468e-15 \\
    20 & 3.149 & (3, 3+1/8)  & 3.232e-17       & 3.873e-17       & 1.050e-16       & 5.005e-16       & 3.959e-16       & \bad 3.337e-02 \\
    21 & 3.149 & (-3, 3+1/8) & 1.043e-16       & 2.178e-17       & 2.989e-15       & 2.642e-15       & 5.205e-16       & 6.698e-15 \\
    22 & 3.277 & (3, 3+1/8)  & 4.345e-16       & 1.803e-16       & 7.954e-16       & 1.264e-15       & 1.397e-14       & \bad 7.796e-01 \\
    23 & 3.277 & (-3, 3+1/8) & 3.241e-16       & 3.454e-17       & 5.567e-16       & 6.517e-16       & 2.187e-16       & 1.612e-13 \\
    24 & 4.000 & (0, 4)      & 1.099e-16       & \bad 7.547e-04  & 3.697e-17       & 5.693e-17       & 5.130e-17       & 3.125e-16 \\
    \bottomrule
  \end{tabular}
  \caption{\M{\phi_\beta}}
  \label{tab:ising_l2}
\end{table}

\begin{table}[htb]
  \centering\small
  \def\M#1{$\mel{\phi_\alpha}{\bar{L}_{+2}}{#1}$}
  \begin{tabular}{*{9}{c}}
    \toprule
    $\ket{\phi_\alpha}$ & $\Delta_\alpha$ & $(s_\alpha, \Delta^{\text{exact}}_\alpha)$ & \M{12} & \M{15} & \M{19} & \M{21} & \M{22} & \M{23} \\
    \midrule
    1  & 0.000 & (0, 0)      & 5.467e-15       & \bad 4.795e-03  & 1.856e-16       & 1.520e-16       & 5.713e-16      & 1.245e-16 \\
    2  & 0.123 & (0, 1/8)    & \good 9.867e-01 & 4.046e-15       & 2.586e-16       & 5.058e-16       & 9.216e-16      & 8.715e-16 \\
    3  & 0.990 & (0, 1)      & 3.940e-15       & \good 9.955e-01 & 1.994e-14       & 1.384e-14       & 1.628e-14      & 1.349e-15 \\
    4  & 1.119 & (1, 1+1/8)  & 6.594e-16       & 2.919e-15       & \good 9.841e-01 & 4.570e-13       & 1.000e-14      & 4.304e-13 \\
    5  & 1.119 & (-1, 1+1/8) & 6.964e-16       & 2.776e-15       & 3.919e-13       & \good 9.842e-01 & 2.353e-12      & \good 9.962e-01 \\
    6  & 2.000 & (1, 2)      & 2.966e-17       & 2.019e-16       & 3.288e-15       & 5.373e-16       & 2.381e-15      & 1.743e-16 \\
    7  & 2.000 & (-1, 2)     & 6.025e-17       & 3.039e-16       & 1.670e-16       & 4.959e-15       & 5.919e-16      & 1.318e-15 \\
    8  & 2.000 & (2, 2)      & 7.485e-17       & 2.251e-16       & 2.593e-16       & 2.083e-16       & 4.436e-16      & 2.246e-16 \\
    9  & 2.000 & (-2, 2)     & 7.495e-17       & 6.037e-16       & 3.122e-16       & 3.257e-16       & 1.903e-15      & 3.502e-16 \\
    10 & 2.114 & (0, 2+1/8)  & \bad 8.935e-04  & 9.438e-16       & 3.131e-16       & 3.338e-16       & 7.598e-16      & 2.725e-16 \\
    11 & 2.154 & (2, 2+1/8)  & 6.331e-16       & 9.592e-17       & 2.916e-16       & 1.403e-16       & 4.221e-16      & 2.540e-16 \\
    12 & 2.154 & (-2, 2+1/8) & 7.313e-16       & 1.787e-16       & 3.496e-16       & 4.773e-16       & 2.210e-15      & 4.931e-16 \\
    13 & 3.010 & (0, 3)      & 1.460e-15       & \bad 1.158e-03  & 6.726e-17       & 5.694e-17       & 1.475e-15      & 8.429e-17 \\
    14 & 3.072 & (2, 3)      & 3.628e-17       & 2.232e-16       & 2.778e-17       & 1.042e-16       & 1.397e-15      & 2.724e-17 \\
    15 & 3.072 & (-2, 3)     & 4.996e-17       & 1.423e-16       & 2.763e-17       & 1.681e-16       & 1.346e-15      & 4.588e-17 \\
    16 & 3.072 & (3, 3)      & 3.907e-17       & 5.139e-16       & 5.574e-17       & 1.623e-16       & 1.521e-15      & 1.383e-16 \\
    17 & 3.072 & (-3, 3)     & 3.601e-17       & 3.837e-16       & 1.870e-16       & 4.195e-16       & 1.606e-14      & 9.056e-17 \\
    18 & 3.149 & (1, 3+1/8)  & 3.890e-16       & 3.388e-17       & \bad 2.615e-03  & 2.466e-15       & 2.468e-15      & 2.766e-17 \\
    19 & 3.149 & (-1, 3+1/8) & 5.707e-16       & 2.357e-17       & 1.371e-15       & \bad 1.438e-02  & 5.981e-15      & \bad 2.638e-04 \\
    20 & 3.149 & (3, 3+1/8)  & 1.043e-16       & 2.178e-17       & 2.989e-15       & 2.642e-15       & 6.698e-15      & 5.205e-16 \\
    21 & 3.149 & (-3, 3+1/8) & 3.232e-17       & 3.873e-17       & 1.050e-16       & 5.005e-16       & \bad 3.337e-02 & 3.959e-16 \\
    22 & 3.277 & (3, 3+1/8)  & 3.241e-16       & 3.454e-17       & 5.567e-16       & 6.517e-16       & 1.612e-13      & 2.187e-16 \\
    23 & 3.277 & (-3, 3+1/8) & 4.345e-16       & 1.803e-16       & 7.954e-16       & 1.264e-15       & \bad 7.796e-01 & 1.397e-14 \\
    24 & 4.000 & (0, 4)      & 1.099e-16       & 7.547e-04       & 3.697e-17       & 5.693e-17       & 3.125e-16      & 5.130e-17 \\
    \bottomrule
  \end{tabular}
  \caption{\M{\phi_\beta}}
  \label{tab:ising_lbar2}
\end{table}

\clearpage

\subsection{Numerical details for the dimer model}

Similar to the Virasoro algebra case, we start by combining dimer model tensor $B^{(0)}$ into $2\times2$ square block to get a new tensor $B=B^{(1)}$ of bond dimension $\chi=4$ and then extract the approximate local current tensor $J$ operator and its holomorphic part $\bar{J}$ from a smaller cylinder of $N=4$ copies of coarse-grained tensor $B$.  $J$ and $\bar{J}$ are then used to construct Kac--Moody generators $J_n$ and $J_{-n}$ on a larger $N=8$ cylinder.

\begin{figure}[htb]
  \centering
  \subfloat[$J_{-1}$]{
    \includegraphics[width=0.32\textwidth]{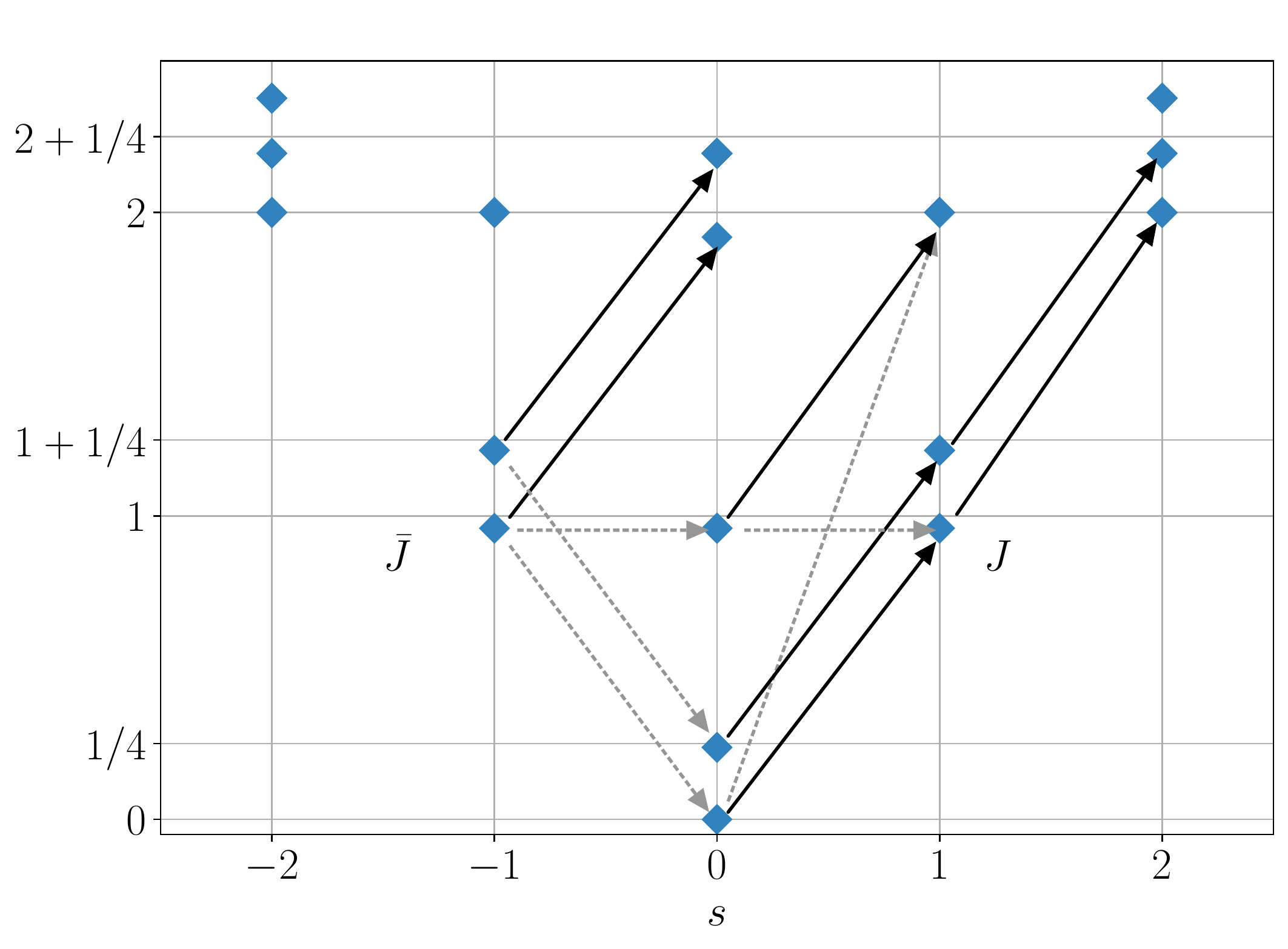}
    \label{fig:dimer_jm1}}
  \qquad
  \subfloat[$\bar{J}_{-1}$]{
    \includegraphics[width=0.32\textwidth]{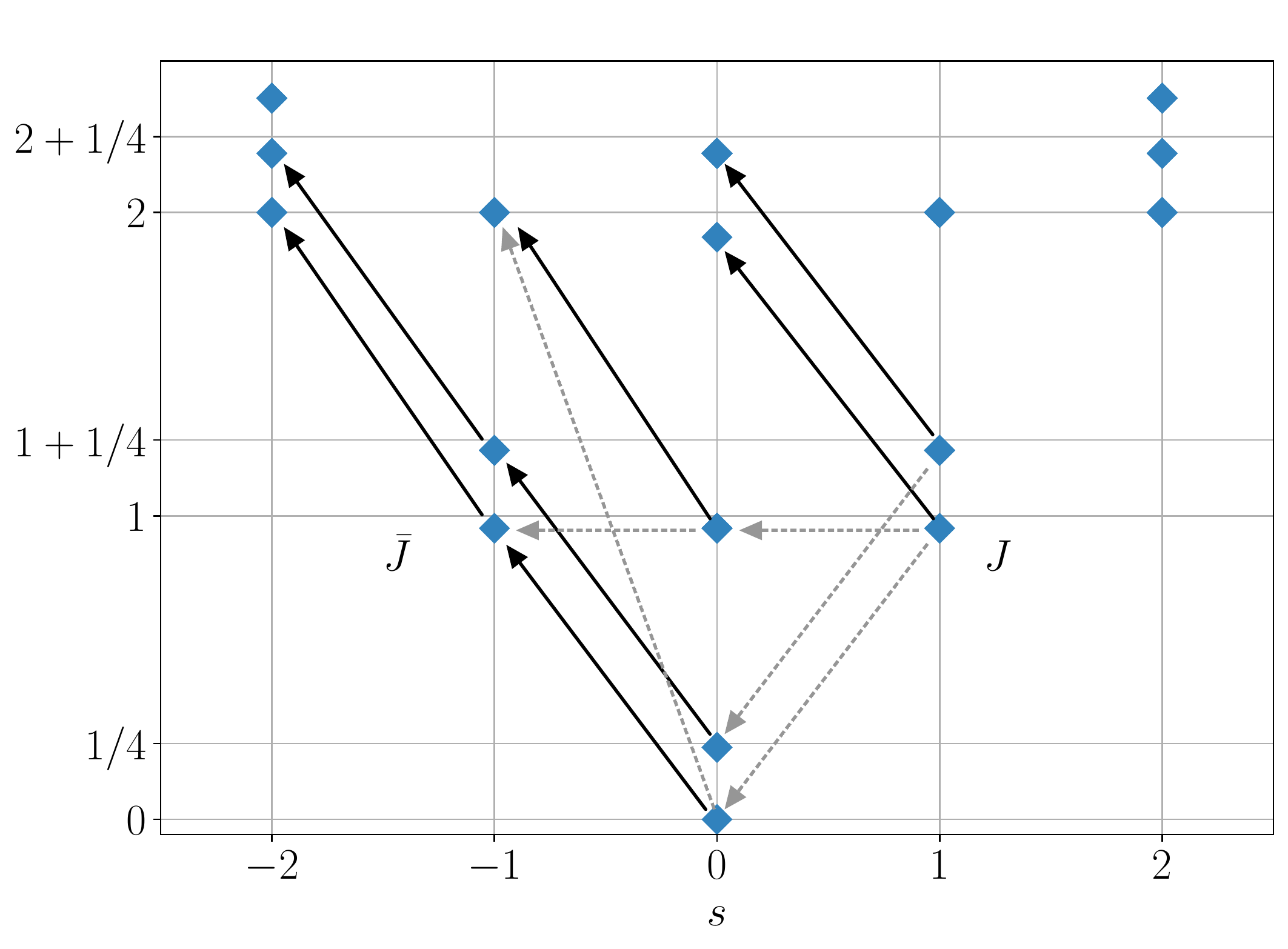}
    \label{fig:dimer_jbarm1}}
  \\
  \subfloat[$J_{-2}$]{
    \includegraphics[width=0.32\textwidth]{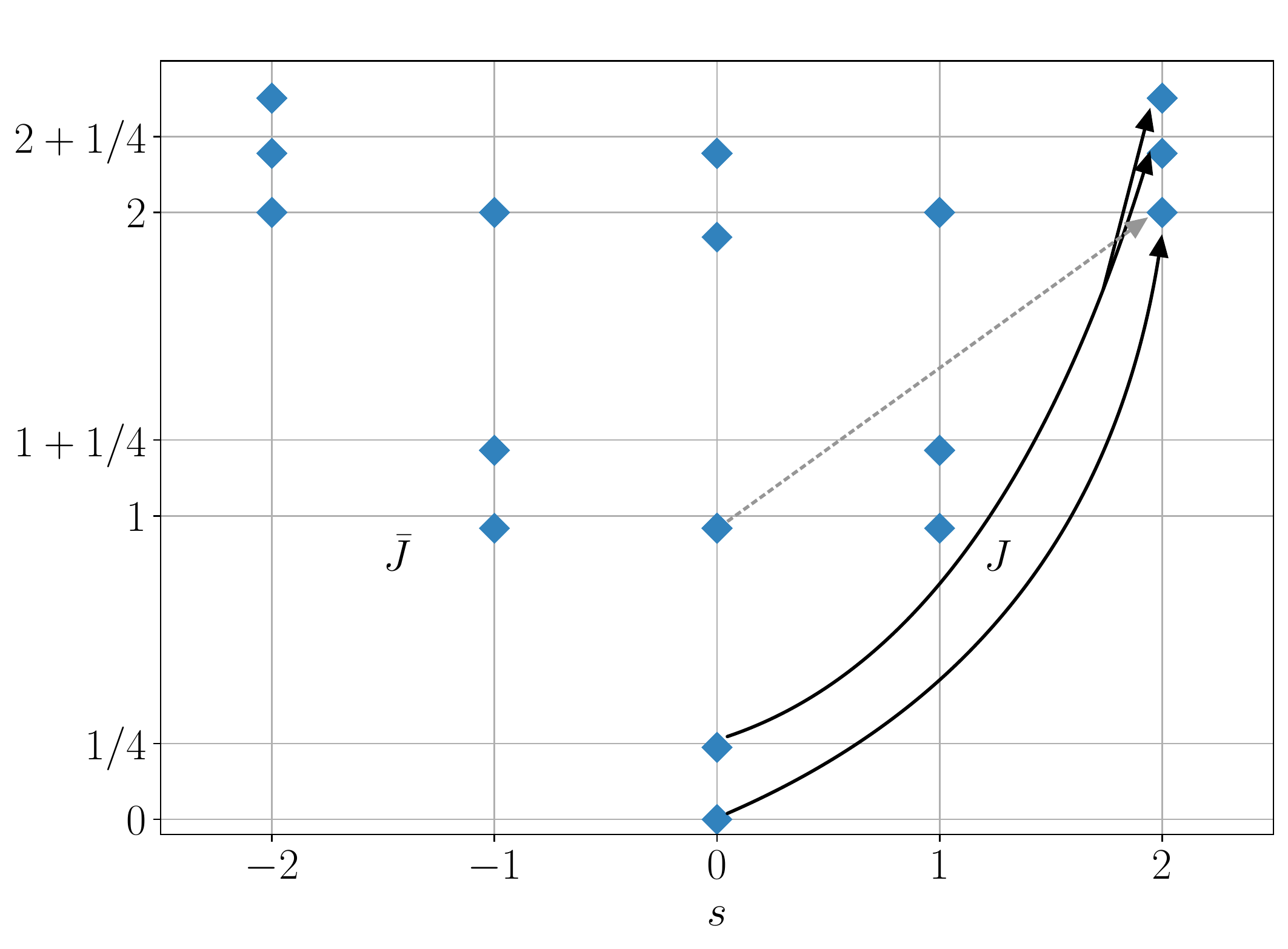}
    \label{fig:dimer_jm2}}
  \qquad
  \subfloat[$\bar{J}_{-2}$]{
    \includegraphics[width=0.32\textwidth]{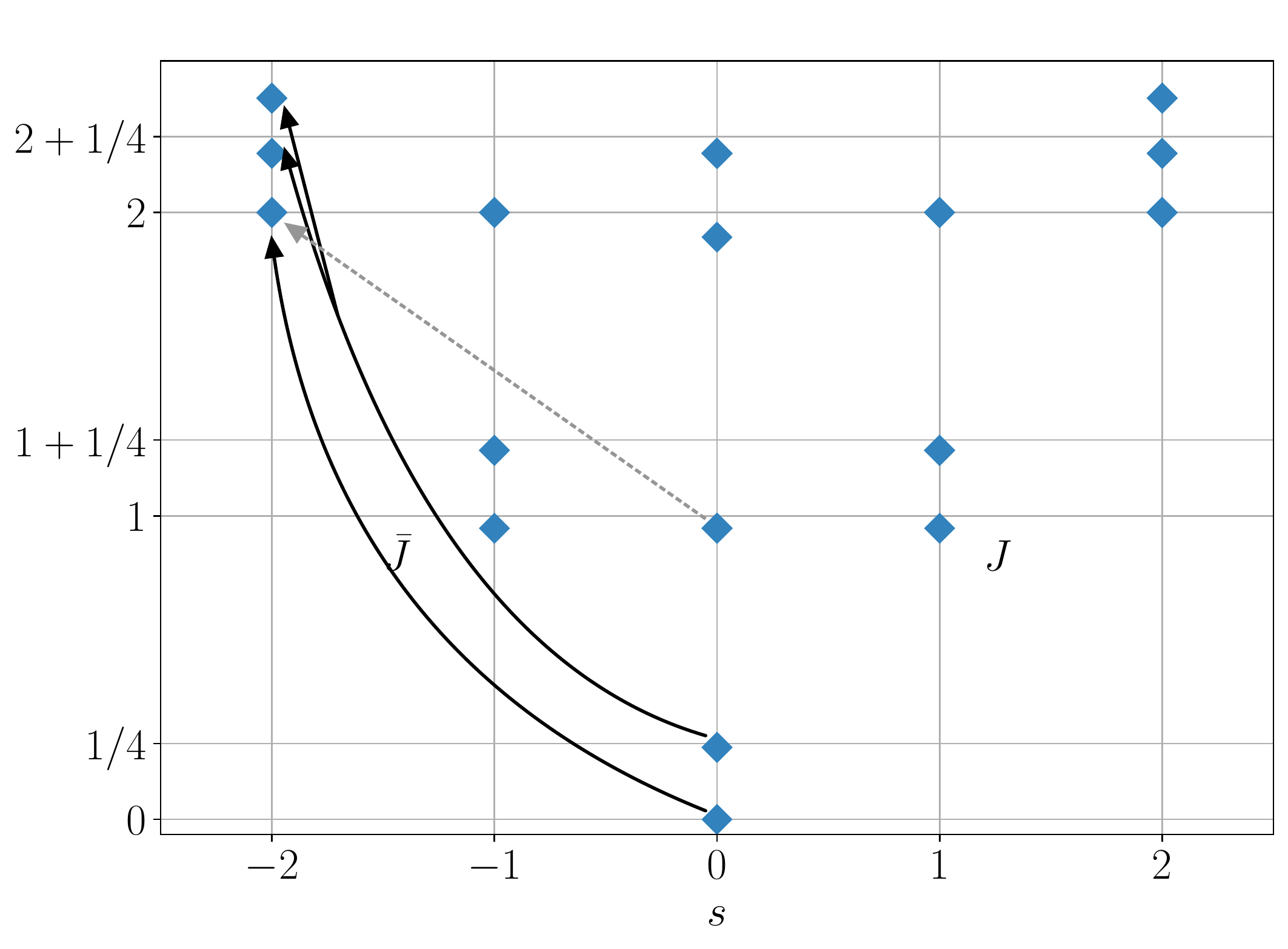}
    \label{fig:dimer_jbarm2}}
  \caption{Actions of lattice Kac--Moody generators $J_{-1, -2}$ and $\bar{J}_{-1, -2}$ in the low energy subspace of transfer matrix consisting of a row of $N=8$ copies of $B$ with bond dimension $\chi=4$. Corresponding matrix elements are listed in Table~\ref{tab:dimer_jm1}--\ref{tab:dimer_jm_2_jbarm2}.}
  \label{fig:j1j2}
\end{figure}

We verify the actions of the lattice Kac--Moody generators by examining the matrix elements $\mel{\phi_\alpha}{J_n}{\phi_\beta}$ and $\mel{\phi_\alpha}{\bar{J}_{n}}{\phi_\beta}$ with $n=-1, -2$ using the low energy eigenstates from a cylinder of $N=8$ sites of $B$. Numerical results of the matrix elements are provided in Table~\ref{tab:dimer_jm1}--\ref{tab:dimer_jm_2_jbarm2} and their behaviors in the low energy subspace are illustrated in Fig.~\ref{fig:j1j2}. The matrix elements between states and their correct Kac--Moody descendant states are highlighted in bold blue in the tables and illustrated correspondingly by solid black arrows in the figures. The erroneous matrix elements are marked in pale blue in the tables and illustrated by dashed gray arrows in the figures.

\begin{table}[htb]
  \centering\small
  \def\M#1{$\mel{\phi_\alpha}{J_{-1}}{#1}$}
  \begin{tabular}{*{10}{c}}
    \toprule
    $\ket{\phi_\alpha}$ & $\Delta_\alpha$ & $(s_\alpha, \Delta^{\text{exact}}_\alpha)$ & \M{1} & \M{2} & \M{4} & \M{8} & \M{9} & \M{10} & \M{14} \\
    \midrule
    1  & 0.000 & (0, 0)      & 3.266e-16       & 9.883e-17       & 6.044e-11       & \bad 2.757e-01  & 9.948e-12       & 2.562e-15       & 2.663e-17 \\
    2  & 0.238 & (0, 1/4)    & 1.064e-16       & 7.638e-16       & 4.130e-16       & 9.487e-15       & 5.270e-16       & \bad 3.124e-01  & 5.666e-11 \\
    3  & 0.238 & (0, 1/4)    & 4.070e-16       & 9.022e-16       & 5.978e-16       & 5.873e-15       & 6.256e-16       & \bad 1.838e-01  & 3.055e-11 \\
    4  & 0.959 & (0, 1)      & 6.138e-11       & 2.088e-15       & 2.491e-11       & \bad 2.632e-01  & 9.498e-12       & 1.912e-15       & 3.097e-17 \\
    5  & 0.959 & (0, 1)      & 1.089e-11       & 5.677e-15       & 1.069e-10       & \bad 4.899e-01  & 1.768e-11       & 5.352e-16       & 7.399e-17 \\
    6  & 0.959 & (0, 1)      & 1.518e-12       & 1.020e-15       & 1.091e-10       & \bad 5.052e-01  & 1.823e-11       & 1.039e-15       & 1.485e-16 \\
    7  & 0.959 & (0, 1)      & 4.075e-12       & 1.820e-15       & 2.025e-10       & \bad 9.031e-01  & 3.259e-11       & 5.185e-16       & 4.227e-17 \\
    8  & 0.959 & (-1, 1)     & 2.689e-11       & 1.859e-15       & 2.990e-11       & 4.527e-11       & 2.670e-16       & 2.239e-16       & 1.371e-16 \\
    9  & 0.959 & (1, 1)      & \good 5.183e-01 & 7.382e-15       & \bad 5.762e-01  & 1.067e-10       & 7.061e-11       & 3.101e-15       & 2.859e-15 \\
    10 & 1.216 & (-1, 1+1/4) & 6.510e-17       & 6.019e-11       & 1.042e-16       & 1.921e-15       & 3.178e-16       & 3.843e-16       & 1.223e-16 \\
    11 & 1.216 & (-1, 1+1/4) & 5.307e-16       & 6.865e-11       & 7.120e-16       & 1.391e-15       & 3.005e-16       & 1.358e-16       & 8.285e-17 \\
    12 & 1.216 & (-1, 1+1/4) & 4.249e-16       & 2.416e-11       & 8.148e-16       & 4.073e-15       & 2.484e-16       & 1.272e-16       & 1.209e-16 \\
    13 & 1.216 & (-1, 1+1/4) & 2.455e-16       & 8.021e-12       & 4.521e-16       & 2.631e-16       & 2.976e-16       & 2.922e-16       & 2.234e-16 \\
    14 & 1.216 & (1, 1+1/4)  & 6.509e-15       & \good 4.781e-01 & 2.564e-15       & 2.758e-15       & 2.587e-16       & 2.079e-16       & 3.778e-16 \\
    15 & 1.216 & (1, 1+1/4)  & 8.550e-15       & \good 7.141e-01 & 3.372e-15       & 1.789e-15       & 2.780e-16       & 1.086e-16       & 3.907e-16 \\
    16 & 1.216 & (1, 1+1/4)  & 4.634e-15       & \good 5.123e-01 & 2.874e-15       & 3.657e-15       & 1.700e-16       & 6.235e-17       & 3.741e-16 \\
    17 & 1.216 & (1, 1+1/4)  & 3.203e-15       & \good 3.976e-01 & 2.022e-15       & 1.913e-15       & 2.805e-16       & 1.757e-16       & 2.675e-16 \\
    18 & 1.919 & (0, 2)      & 2.812e-16       & 4.137e-16       & 7.801e-11       & \good 3.559e-01 & 1.284e-11       & 7.785e-16       & 9.478e-17 \\
    19 & 2.000 & (-1, 2)     & 1.968e-12       & 6.976e-16       & 9.505e-13       & 2.307e-16       & 4.849e-12       & 4.178e-15       & 2.715e-15 \\
    20 & 2.000 & (-1, 2)     & 8.242e-12       & 8.128e-17       & 4.844e-12       & 3.549e-17       & 3.881e-12       & 7.392e-16       & 6.227e-16 \\
    21 & 2.000 & (-1, 2)     & 1.450e-11       & 4.031e-17       & 1.368e-11       & 1.027e-16       & 7.315e-12       & 5.798e-16       & 1.136e-15 \\
    22 & 2.000 & (-1, 2)     & 1.831e-11       & 8.237e-17       & 1.993e-12       & 2.046e-16       & 5.108e-12       & 9.647e-16       & 1.998e-16 \\
    23 & 2.000 & (1, 2)      & \bad 4.648e-01  & 3.917e-15       & \good 3.806e-01 & 6.492e-11       & 4.055e-11       & 2.128e-16       & 1.392e-15 \\
    24 & 2.000 & (1, 2)      & \bad 1.437e-01  & 1.826e-15       & \good 6.135e-01 & 1.064e-10       & 5.895e-11       & 1.785e-15       & 9.359e-16 \\
    25 & 2.000 & (1, 2)      & \bad 3.164e-01  & 3.711e-15       & \good 4.870e-01 & 7.362e-11       & 2.781e-11       & 9.552e-16       & 2.343e-15 \\
    26 & 2.000 & (1, 2)      & \bad 2.899e-01  & 3.117e-15       & \good 1.268e-01 & 1.182e-11       & 2.874e-12       & 2.934e-15       & 1.407e-15 \\
    27 & 2.000 & (-2, 2)     & 8.537e-12       & 4.170e-16       & 3.216e-12       & 4.579e-16       & 5.014e-12       & 1.917e-15       & 2.484e-15 \\
    28 & 2.000 & (-2, 2)     & 1.645e-11       & 1.588e-16       & 1.545e-11       & 3.429e-16       & 1.537e-11       & 6.567e-16       & 8.569e-16 \\
    29 & 2.000 & (-2, 2)     & 2.164e-11       & 1.943e-16       & 2.632e-11       & 4.852e-17       & 1.764e-11       & 2.056e-16       & 1.091e-15 \\
    30 & 2.000 & (-2, 2)     & 1.067e-11       & 2.014e-17       & 2.818e-12       & 4.102e-16       & 1.128e-12       & 4.001e-16       & 4.967e-16 \\
    31 & 2.000 & (2, 2)      & 1.044e-11       & 2.259e-16       & 7.561e-11       & 2.277e-11       & \good 3.689e-01 & 1.056e-15       & 4.923e-16 \\
    32 & 2.000 & (2, 2)      & 2.492e-11       & 7.301e-17       & 6.679e-11       & 2.410e-11       & \good 3.904e-01 & 6.301e-16       & 2.271e-15 \\
    33 & 2.000 & (2, 2)      & 1.134e-11       & 2.059e-16       & 3.484e-11       & 4.293e-12       & \good 6.957e-02 & 9.089e-16       & 2.796e-15 \\
    34 & 2.000 & (2, 2)      & 3.939e-12       & 6.725e-17       & 5.227e-11       & 1.975e-11       & \good 3.200e-01 & 2.253e-16       & 2.540e-15 \\
    35 & 2.195 & (0, 2+1/4)  & 2.053e-15       & 2.327e-16       & 1.979e-15       & 1.877e-15       & 1.368e-15       & \good 3.696e-01 & 2.602e-12 \\
    36 & 2.195 & (0, 2+1/4)  & 6.288e-16       & 3.234e-16       & 8.580e-16       & 1.791e-15       & 3.288e-16       & \good 5.306e-02 & 1.510e-11 \\
    37 & 2.195 & (0, 2+1/4)  & 1.745e-15       & 3.589e-16       & 1.138e-15       & 3.367e-15       & 1.055e-15       & \good 2.347e-01 & 7.151e-12 \\
    38 & 2.195 & (0, 2+1/4)  & 1.299e-15       & 1.713e-16       & 1.168e-15       & 9.117e-16       & 5.968e-16       & \good 1.773e-01 & 7.317e-12 \\
    39 & 2.195 & (0, 2+1/4)  & 1.242e-15       & 1.738e-16       & 3.561e-16       & 6.054e-16       & 5.048e-16       & \good 9.924e-02 & 8.668e-12 \\
    40 & 2.195 & (0, 2+1/4)  & 3.270e-16       & 8.875e-17       & 5.018e-16       & 3.345e-16       & 2.662e-16       & \good 1.430e-01 & 1.187e-11 \\
    41 & 2.195 & (0, 2+1/4)  & 3.796e-17       & 1.624e-16       & 1.758e-16       & 1.203e-15       & 1.682e-16       & \good 3.495e-01 & 1.165e-11 \\
    42 & 2.195 & (0, 2+1/4)  & 2.522e-17       & 3.743e-16       & 9.655e-17       & 6.641e-16       & 1.456e-16       & \good 9.315e-02 & 2.102e-11 \\
    43 & 2.195 & (-2, 2+1/4) & 1.138e-15       & 5.302e-16       & 8.395e-16       & 3.795e-16       & 6.918e-16       & 2.725e-12       & 2.541e-12 \\
    44 & 2.195 & (-2, 2+1/4) & 1.235e-15       & 3.312e-16       & 1.542e-15       & 3.598e-17       & 5.377e-16       & 1.180e-12       & 3.255e-12 \\
    45 & 2.195 & (2, 2+1/4)  & 1.489e-15       & 1.485e-16       & 1.112e-15       & 7.241e-16       & 2.307e-15       & 2.999e-11       & \good  4.081e-01 \\
    46 & 2.195 & (2, 2+1/4)  & 1.675e-15       & 7.517e-17       & 1.038e-15       & 6.622e-16       & 8.062e-16       & 2.772e-11       & \good  1.787e-01 \\
    47 & 2.377 & (-2, 2+1/4) & 4.418e-16       & 3.444e-16       & 3.947e-16       & 6.947e-17       & 1.292e-16       & 1.086e-15       & 3.550e-12 \\
    48 & 2.377 & (-2, 2+1/4) & 1.031e-15       & 1.532e-16       & 1.239e-15       & 4.650e-16       & 4.607e-16       & 8.862e-16       & 2.474e-12 \\
    49 & 2.377 & (-2, 2+1/4) & 6.110e-16       & 9.260e-17       & 2.389e-16       & 8.948e-17       & 9.831e-16       & 1.028e-15       & 5.179e-12 \\
    50 & 2.377 & (-2, 2+1/4) & 1.323e-15       & 1.048e-17       & 4.112e-16       & 4.093e-17       & 7.262e-16       & 1.550e-15       & 2.151e-12 \\
    51 & 2.377 & (2, 2+1/4)  & 1.052e-15       & 2.213e-16       & 1.268e-15       & 9.389e-16       & 2.294e-15       & 5.728e-11       & \good 3.248e-01 \\
    52 & 2.377 & (2, 2+1/4)  & 6.045e-16       & 8.078e-17       & 1.039e-15       & 9.054e-16       & 6.901e-16       & 2.348e-11       & \good 2.142e-01 \\
    53 & 2.377 & (2, 2+1/4)  & 8.075e-16       & 1.878e-16       & 2.008e-15       & 5.722e-16       & 1.860e-15       & 2.249e-11       & \good 1.586e-01 \\
    54 & 2.377 & (2, 2+1/4)  & 2.023e-15       & 2.498e-16       & 9.265e-16       & 4.246e-16       & 9.182e-16       & 1.254e-11       & \good 1.564e-01 \\
    \bottomrule
  \end{tabular}
  \caption{\M{\phi_\beta}}
  \label{tab:dimer_jm1}
\end{table}

\begin{table}[htb]
  \centering\small
  \def\M#1{$\mel{\phi_\alpha}{\bar{J}_{-1}}{#1}$}
  \begin{tabular}{*{10}{c}}
    \toprule
    $\ket{\phi_\alpha}$ & $\Delta_\alpha$ & $(s_\alpha, \Delta^{\text{exact}}_\alpha)$ & \M{1} & \M{2} & \M{4} & \M{8} & \M{9} & \M{10} & \M{14} \\
    \midrule
    1  & 0.000 & (0, 0)      & 3.020e-16       & 5.799e-17       & 3.742e-11       & 1.430e-11       & \bad 2.318e-01  & 2.456e-16       & 1.159e-14 \\
    2  & 0.238 & (0, 1/4)    & 1.135e-16       & 1.266e-16       & 5.385e-16       & 5.564e-16       & 7.680e-15       & 3.914e-11       & \bad 4.483e-01 \\
    3  & 0.238 & (0, 1/4)    & 3.490e-16       & 5.037e-16       & 1.165e-15       & 1.463e-15       & 4.741e-15       & 7.515e-11       & \bad 5.998e-01 \\
    4  & 0.959 & (0, 1)      & 9.915e-11       & 3.506e-16       & 9.197e-11       & 1.378e-11       & \bad 2.232e-01  & 1.913e-16       & 1.682e-15 \\
    5  & 0.959 & (0, 1)      & 1.805e-11       & 4.414e-15       & 9.393e-11       & 3.324e-11       & \bad 5.386e-01  & 4.190e-16       & 4.006e-15 \\
    6  & 0.959 & (0, 1)      & 2.372e-12       & 1.384e-15       & 9.074e-11       & 3.556e-11       & \bad 5.761e-01  & 3.588e-16       & 3.894e-15 \\
    7  & 0.959 & (0, 1)      & 6.623e-12       & 1.630e-15       & 1.102e-10       & 3.932e-11       & \bad 6.371e-01  & 3.628e-16       & 3.257e-15 \\
    8  & 0.959 & (-1, 1)     & \good 5.183e-01 & 3.244e-15       & \bad 6.618e-01  & 1.461e-10       & 5.360e-11       & 3.492e-15       & 3.110e-15 \\
    9  & 0.959 & (1, 1)      & 2.224e-11       & 7.495e-16       & 2.840e-11       & 5.159e-16       & 1.825e-11       & 1.598e-16       & 4.967e-16 \\
    10 & 1.216 & (-1, 1+1/4) & 4.985e-15       & \good 3.214e-01 & 1.872e-15       & 7.236e-16       & 3.155e-15       & 1.660e-16       & 7.968e-17 \\
    11 & 1.216 & (-1, 1+1/4) & 2.734e-15       & \good 5.641e-01 & 2.131e-15       & 5.335e-16       & 2.215e-15       & 7.069e-16       & 1.585e-16 \\
    12 & 1.216 & (-1, 1+1/4) & 2.698e-15       & \good 3.587e-01 & 8.678e-16       & 6.764e-16       & 3.653e-15       & 5.304e-16       & 3.257e-16 \\
    13 & 1.216 & (-1, 1+1/4) & 9.450e-15       & \good 3.931e-01 & 4.147e-15       & 1.293e-16       & 1.070e-15       & 4.191e-16       & 9.056e-17 \\
    14 & 1.216 & (1, 1+1/4)  & 5.947e-18       & 4.593e-11       & 1.085e-16       & 3.680e-16       & 3.192e-15       & 1.567e-16       & 3.594e-17 \\
    15 & 1.216 & (1, 1+1/4)  & 7.334e-17       & 4.432e-11       & 3.793e-17       & 6.033e-16       & 1.814e-15       & 9.617e-17       & 1.092e-16 \\
    16 & 1.216 & (1, 1+1/4)  & 1.002e-16       & 1.975e-12       & 1.814e-16       & 4.179e-16       & 3.739e-15       & 1.424e-16       & 1.719e-16 \\
    17 & 1.216 & (1, 1+1/4)  & 2.789e-16       & 1.254e-11       & 3.339e-17       & 7.064e-16       & 1.273e-15       & 5.418e-17       & 2.423e-16 \\
    18 & 1.919 & (0, 2)      & 1.419e-16       & 6.304e-16       & 4.830e-11       & 1.846e-11       & \good 2.992e-01 & 4.875e-16       & 1.008e-15 \\
    19 & 2.000 & (-1, 2)     & \bad 3.999e-01  & 3.270e-15       & \good 3.811e-02 & 3.825e-11       & 9.864e-12       & 1.938e-15       & 3.278e-15 \\
    20 & 2.000 & (-1, 2)     & \bad 1.835e-01  & 2.070e-15       & \good 2.333e-01 & 3.843e-11       & 2.003e-11       & 3.927e-15       & 4.149e-15 \\
    21 & 2.000 & (-1, 2)     & \bad 2.438e-01  & 1.934e-15       & \good 2.347e-01 & 2.554e-11       & 1.580e-11       & 2.017e-15       & 1.642e-15 \\
    22 & 2.000 & (-1, 2)     & \bad 4.389e-01  & 5.504e-15       & \good 1.047e-01 & 4.353e-11       & 1.976e-11       & 2.228e-15       & 3.253e-15 \\
    23 & 2.000 & (1, 2)      & 1.002e-11       & 2.479e-17       & 5.180e-12       & 5.637e-12       & 1.646e-16       & 9.017e-16       & 1.216e-15 \\
    24 & 2.000 & (1, 2)      & 2.807e-12       & 2.346e-16       & 2.478e-12       & 1.427e-11       & 8.927e-17       & 7.120e-16       & 2.025e-15 \\
    25 & 2.000 & (1, 2)      & 1.079e-11       & 3.719e-16       & 6.176e-12       & 5.451e-12       & 2.984e-17       & 1.597e-15       & 9.887e-16 \\
    26 & 2.000 & (1, 2)      & 2.358e-11       & 8.030e-17       & 7.652e-12       & 6.347e-12       & 5.828e-16       & 1.393e-15       & 1.055e-15 \\
    27 & 2.000 & (-2, 2)     & 1.140e-11       & 5.040e-16       & 1.463e-10       & \good 6.736e-01 & 2.431e-11       & 2.067e-15       & 1.468e-15 \\
    28 & 2.000 & (-2, 2)     & 1.313e-11       & 1.233e-16       & 6.710e-11       & \good 3.379e-01 & 1.219e-11       & 1.927e-15       & 5.313e-16 \\
    29 & 2.000 & (-2, 2)     & 1.953e-11       & 2.725e-16       & 6.582e-12       & \good 5.926e-02 & 2.138e-12       & 1.027e-15       & 2.286e-16 \\
    30 & 2.000 & (-2, 2)     & 3.216e-11       & 4.221e-17       & 2.207e-11       & \good 1.419e-01 & 5.118e-12       & 2.223e-15       & 2.871e-16 \\
    31 & 2.000 & (2, 2)      & 2.261e-12       & 3.761e-16       & 1.231e-12       & 1.064e-11       & 5.813e-16       & 9.305e-16       & 9.373e-16 \\
    32 & 2.000 & (2, 2)      & 3.628e-13       & 1.581e-16       & 4.027e-12       & 3.708e-12       & 2.810e-16       & 1.290e-15       & 5.026e-16 \\
    33 & 2.000 & (2, 2)      & 5.142e-12       & 1.018e-17       & 2.166e-12       & 1.184e-12       & 3.061e-16       & 2.051e-15       & 6.268e-16 \\
    34 & 2.000 & (2, 2)      & 6.406e-12       & 2.737e-17       & 1.730e-12       & 7.993e-13       & 1.112e-16       & 2.069e-15       & 3.574e-16 \\
    35 & 2.195 & (0, 2+1/4)  & 2.271e-15       & 3.340e-16       & 1.223e-15       & 3.944e-15       & 1.028e-15       & 9.913e-12       & \good 2.836e-01 \\
    36 & 2.195 & (0, 2+1/4)  & 1.209e-15       & 3.093e-16       & 8.933e-16       & 1.293e-15       & 8.158e-16       & 2.570e-11       & \good 1.279e-01 \\
    37 & 2.195 & (0, 2+1/4)  & 3.761e-16       & 1.995e-16       & 2.839e-16       & 1.739e-15       & 1.743e-15       & 1.777e-11       & \good 1.523e-01 \\
    38 & 2.195 & (0, 2+1/4)  & 7.344e-17       & 1.465e-16       & 1.024e-15       & 7.608e-16       & 1.464e-15       & 1.269e-11       & \good 2.161e-01 \\
    39 & 2.195 & (0, 2+1/4)  & 1.450e-15       & 1.452e-16       & 9.658e-16       & 7.077e-16       & 1.262e-15       & 1.628e-11       & \good 1.951e-01 \\
    40 & 2.195 & (0, 2+1/4)  & 1.645e-15       & 1.447e-16       & 1.149e-15       & 4.737e-16       & 7.450e-16       & 9.775e-12       & \good 7.257e-02 \\
    41 & 2.195 & (0, 2+1/4)  & 2.640e-16       & 3.336e-16       & 5.001e-16       & 6.720e-16       & 1.304e-15       & 1.739e-11       & \good 2.373e-01 \\
    42 & 2.195 & (0, 2+1/4)  & 1.071e-16       & 2.901e-16       & 4.304e-16       & 2.360e-16       & 5.700e-16       & 2.373e-11       & \good 2.880e-01 \\
    43 & 2.195 & (-2, 2+1/4) & 1.334e-15       & 2.948e-16       & 1.152e-15       & 2.702e-15       & 3.894e-18       & \good 3.877e-01 & 1.518e-11 \\
    44 & 2.195 & (-2, 2+1/4) & 2.202e-15       & 2.793e-16       & 7.729e-16       & 4.107e-15       & 4.027e-16       & \good 1.493e-01 & 1.477e-11 \\
    45 & 2.195 & (2, 2+1/4)  & 1.482e-15       & 4.563e-16       & 4.938e-16       & 1.865e-15       & 3.565e-17       & 1.042e-12       & 4.851e-13 \\
    46 & 2.195 & (2, 2+1/4)  & 4.905e-16       & 2.675e-16       & 6.223e-16       & 9.047e-16       & 3.614e-17       & 5.295e-12       & 5.308e-16 \\
    47 & 2.377 & (-2, 2+1/4) & 5.851e-16       & 3.386e-17       & 3.655e-16       & 1.372e-15       & 2.725e-16       & \good 2.010e-01 & 2.553e-11 \\
    48 & 2.377 & (-2, 2+1/4) & 5.408e-16       & 1.632e-16       & 6.563e-17       & 1.133e-15       & 6.621e-16       & \good 5.170e-01 & 4.421e-11 \\
    49 & 2.377 & (-2, 2+1/4) & 7.664e-16       & 5.823e-17       & 4.148e-16       & 2.253e-15       & 2.173e-16       & \good 3.372e-01 & 2.652e-11 \\
    50 & 2.377 & (-2, 2+1/4) & 7.657e-16       & 1.805e-16       & 2.843e-16       & 1.822e-15       & 3.796e-16       & \good 3.870e-01 & 4.651e-12 \\
    51 & 2.377 & (2, 2+1/4)  & 2.777e-15       & 2.141e-16       & 1.112e-15       & 1.403e-15       & 3.865e-16       & 2.678e-12       & 2.201e-12 \\
    52 & 2.377 & (2, 2+1/4)  & 8.471e-16       & 4.432e-16       & 3.088e-16       & 4.609e-16       & 4.860e-17       & 8.292e-12       & 2.122e-16 \\
    53 & 2.377 & (2, 2+1/4)  & 1.658e-15       & 6.381e-17       & 9.357e-17       & 5.356e-16       & 1.621e-16       & 7.703e-12       & 6.784e-17 \\
    54 & 2.377 & (2, 2+1/4)  & 9.213e-16       & 1.814e-16       & 3.439e-16       & 3.685e-16       & 2.471e-17       & 1.386e-12       & 1.645e-15 \\
    \bottomrule
  \end{tabular}
  \caption{\M{\phi_\beta}}
  \label{tab:dimer_jbarm1}
\end{table}

\begin{table}[htb]
  \centering\small
  \def\M#1{$\mel{\phi_\alpha}{J_{-2}}{#1}$}
  \def\Mb#1{$\mel{\phi_\alpha}{\bar{J}_{-2}}{#1}$}
  \begin{tabular}{*{10}{c}}
    \toprule
    $\ket{\phi_\alpha}$ & $\Delta_\alpha$ & $(s_\alpha, \Delta^{\text{exact}}_\alpha)$ & \M1 & \Mb1 & \M2 & \Mb2 & \M4 & \Mb4 \\
    \midrule
    1  & 0.000 & (0, 0)      & 2.194e-16       & 3.659e-16       & 1.213e-17       & 1.032e-17       & 2.798e-16      & 2.106e-16 \\
    2  & 0.238 & (0, 1/4)    & 8.377e-19       & 4.856e-17       & 1.364e-16       & 4.728e-16       & 9.702e-17      & 5.544e-17 \\
    3  & 0.238 & (0, 1/4)    & 4.599e-17       & 1.376e-16       & 3.707e-16       & 3.182e-16       & 1.963e-16      & 1.692e-16 \\
    4  & 0.959 & (0, 1)      & 9.406e-17       & 1.472e-16       & 2.741e-17       & 7.702e-17       & 3.837e-16      & 1.160e-16 \\
    5  & 0.959 & (0, 1)      & 1.101e-16       & 2.293e-16       & 7.341e-17       & 5.621e-17       & 7.833e-16      & 3.063e-16 \\
    6  & 0.959 & (0, 1)      & 3.184e-16       & 3.024e-16       & 2.215e-16       & 1.385e-16       & 6.661e-16      & 3.529e-17 \\
    7  & 0.959 & (0, 1)      & 1.401e-16       & 5.457e-16       & 1.898e-16       & 1.231e-16       & 7.434e-17      & 2.547e-16 \\
    8  & 0.959 & (-1, 1)     & 5.993e-17       & 1.192e-16       & 3.856e-17       & 1.080e-16       & 1.575e-16      & 2.826e-12 \\
    9  & 0.959 & (1, 1)      & 1.117e-16       & 1.813e-16       & 1.013e-17       & 1.377e-17       & 4.565e-12      & 1.624e-16 \\
    10 & 1.216 & (-1, 1+1/4) & 9.443e-17       & 1.048e-17       & 2.012e-16       & 2.514e-16       & 2.827e-16      & 8.619e-16 \\
    11 & 1.216 & (-1, 1+1/4) & 7.415e-17       & 3.687e-17       & 2.573e-16       & 1.988e-16       & 3.123e-16      & 7.400e-16 \\
    12 & 1.216 & (-1, 1+1/4) & 3.593e-17       & 1.049e-16       & 2.114e-16       & 4.771e-16       & 3.535e-16      & 1.078e-15 \\
    13 & 1.216 & (-1, 1+1/4) & 2.256e-16       & 1.911e-16       & 2.907e-16       & 5.072e-16       & 3.243e-16      & 1.104e-15 \\
    14 & 1.216 & (1, 1+1/4)  & 4.307e-17       & 1.350e-16       & 1.811e-16       & 2.119e-16       & 3.112e-16      & 2.002e-16 \\
    15 & 1.216 & (1, 1+1/4)  & 7.404e-17       & 2.228e-16       & 1.634e-16       & 2.107e-16       & 3.727e-16      & 2.639e-16 \\
    16 & 1.216 & (1, 1+1/4)  & 4.849e-17       & 1.392e-16       & 3.222e-16       & 3.846e-16       & 2.661e-16      & 1.687e-16 \\
    17 & 1.216 & (1, 1+1/4)  & 1.340e-16       & 1.475e-16       & 2.968e-16       & 2.126e-16       & 4.533e-16      & 2.898e-16 \\
    18 & 1.919 & (0, 2)      & 1.844e-16       & 2.281e-16       & 1.862e-16       & 2.694e-16       & 3.284e-16      & 1.916e-16 \\
    19 & 2.000 & (-1, 2)     & 1.502e-12       & 5.731e-12       & 1.147e-15       & 2.872e-15       & 4.912e-12      & 6.888e-11 \\
    20 & 2.000 & (-1, 2)     & 4.641e-12       & 1.237e-11       & 9.452e-16       & 1.328e-15       & 4.387e-12      & 3.122e-11 \\
    21 & 2.000 & (-1, 2)     & 5.865e-12       & 9.658e-12       & 1.714e-15       & 1.891e-15       & 5.425e-12      & 8.727e-11 \\
    22 & 2.000 & (-1, 2)     & 6.081e-12       & 4.235e-11       & 7.000e-16       & 1.299e-15       & 7.953e-12      & 1.034e-11 \\
    23 & 2.000 & (1, 2)      & 3.906e-12       & 4.488e-12       & 2.224e-15       & 9.620e-16       & 6.344e-11      & 5.522e-12 \\
    24 & 2.000 & (1, 2)      & 2.882e-11       & 8.244e-12       & 1.473e-15       & 5.436e-16       & 1.373e-10      & 5.865e-12 \\
    25 & 2.000 & (1, 2)      & 1.622e-11       & 9.787e-12       & 1.425e-15       & 1.356e-15       & 7.522e-11      & 2.447e-11 \\
    26 & 2.000 & (1, 2)      & 7.050e-12       & 1.439e-11       & 1.606e-15       & 8.611e-16       & 6.315e-11      & 1.006e-11 \\
    27 & 2.000 & (-2, 2)     & 7.211e-12       & \good 3.044e-01 & 1.727e-15       & 2.134e-15       & 5.893e-12      & \bad 5.222e-01 \\
    28 & 2.000 & (-2, 2)     & 1.706e-11       & \good 4.629e-02 & 1.484e-15       & 5.751e-15       & 1.096e-11      & \bad 4.649e-01 \\
    29 & 2.000 & (-2, 2)     & 1.511e-11       & \good 2.115e-01 & 1.451e-15       & 2.826e-15       & 1.075e-11      & \bad 3.456e-01 \\
    30 & 2.000 & (-2, 2)     & 3.824e-12       & \good 6.287e-01 & 7.909e-16       & 9.918e-16       & 3.807e-12      & \bad 2.923e-01 \\
    31 & 2.000 & (2, 2)      & \good 3.380e-01 & 6.555e-12       & 7.619e-16       & 1.541e-15       & \bad 2.352e-01 & 1.286e-11 \\
    32 & 2.000 & (2, 2)      & \good 2.816e-01 & 8.877e-12       & 3.110e-15       & 1.371e-15       & \bad 1.694e-01 & 1.257e-11 \\
    33 & 2.000 & (2, 2)      & \good 2.577e-01 & 1.012e-11       & 4.569e-15       & 9.969e-16       & \bad 2.264e-01 & 2.027e-11 \\
    34 & 2.000 & (2, 2)      & \good 4.637e-01 & 3.614e-12       & 2.901e-15       & 2.480e-15       & \bad 5.672e-01 & 2.018e-12 \\
    35 & 2.195 & (0, 2+1/4)  & 1.167e-15       & 1.828e-15       & 9.615e-13       & 2.382e-12       & 2.581e-15      & 2.690e-15 \\
    36 & 2.195 & (0, 2+1/4)  & 1.305e-15       & 6.678e-16       & 2.791e-13       & 1.038e-12       & 1.031e-15      & 1.240e-15 \\
    37 & 2.195 & (0, 2+1/4)  & 1.047e-15       & 1.135e-15       & 1.271e-12       & 2.231e-12       & 1.554e-15      & 1.661e-15 \\
    38 & 2.195 & (0, 2+1/4)  & 5.541e-16       & 1.705e-16       & 6.295e-13       & 6.343e-13       & 4.878e-16      & 7.357e-16 \\
    39 & 2.195 & (0, 2+1/4)  & 7.794e-16       & 3.391e-16       & 5.345e-13       & 3.914e-13       & 1.329e-16      & 2.473e-16 \\
    40 & 2.195 & (0, 2+1/4)  & 5.086e-16       & 9.840e-16       & 1.668e-13       & 1.307e-13       & 1.170e-15      & 3.583e-16 \\
    41 & 2.195 & (0, 2+1/4)  & 1.193e-16       & 2.020e-16       & 1.325e-13       & 1.266e-13       & 1.753e-16      & 2.733e-16 \\
    42 & 2.195 & (0, 2+1/4)  & 6.604e-17       & 5.515e-17       & 1.889e-14       & 5.218e-14       & 1.103e-16      & 3.927e-16 \\
    43 & 2.195 & (-2, 2+1/4) & 1.019e-15       & 1.238e-15       & 3.150e-12       & \good 4.075e-01 & 9.162e-16      & 2.515e-15 \\
    44 & 2.195 & (-2, 2+1/4) & 4.996e-16       & 1.763e-15       & 2.271e-12       & \good 3.051e-01 & 4.302e-16      & 3.554e-15 \\
    45 & 2.195 & (2, 2+1/4)  & 1.530e-15       & 8.424e-16       & \good 3.537e-01 & 2.683e-12       & 5.932e-16      & 1.291e-15 \\
    46 & 2.195 & (2, 2+1/4)  & 7.470e-16       & 7.818e-16       & \good 3.272e-01 & 2.380e-12       & 3.184e-15      & 1.618e-15 \\
    47 & 2.377 & (-2, 2+1/4) & 2.677e-16       & 5.734e-15       & 7.256e-12       & \good 2.523e-01 & 3.928e-16      & 9.787e-16 \\
    48 & 2.377 & (-2, 2+1/4) & 1.177e-15       & 7.462e-15       & 2.931e-12       & \good 1.915e-01 & 1.809e-16      & 6.705e-16 \\
    49 & 2.377 & (-2, 2+1/4) & 1.971e-15       & 5.454e-15       & 3.323e-12       & \good 4.874e-01 & 1.894e-15      & 1.455e-15 \\
    50 & 2.377 & (-2, 2+1/4) & 1.225e-15       & 6.717e-15       & 2.831e-12       & \good 5.481e-01 & 2.683e-16      & 7.247e-16 \\
    51 & 2.377 & (2, 2+1/4)  & 5.192e-15       & 6.196e-16       & \good 5.129e-01 & 2.238e-12       & 1.318e-15      & 6.241e-16 \\
    52 & 2.377 & (2, 2+1/4)  & 7.118e-16       & 1.705e-15       & \good 1.430e-01 & 1.313e-11       & 1.920e-15      & 5.857e-16 \\
    53 & 2.377 & (2, 2+1/4)  & 4.160e-15       & 1.017e-15       & \good 8.675e-02 & 2.092e-12       & 3.247e-15      & 1.300e-15 \\
    54 & 2.377 & (2, 2+1/4)  & 8.380e-15       & 7.529e-16       & \good 7.333e-01 & 1.266e-12       & 1.726e-15      & 3.195e-16 \\
    \bottomrule
  \end{tabular}
  \caption{\M{\phi_\beta} and \Mb{\phi_\beta}}
  \label{tab:dimer_jm_2_jbarm2}
\end{table}

\end{document}